\documentclass[twocolumn]{aastex631}
\usepackage{amsmath}
\usepackage{graphicx}
\usepackage{subfigure}
\usepackage{booktabs}

\usepackage{xcolor}

\received{September 15, 2024}
\revised{November 14, 2024}
\accepted{November 15, 2024}

\shorttitle{Stellar Collision}

\shortauthors{Yu and Lai}

\graphicspath{{./}{figures/}}

\begin{document}
\title{Binary Stars Approaching Supermassive Black Holes: Tidal Break-up, Double Stellar Disruptions and Stellar Collision}
\correspondingauthor{Fangyuan Yu, Dong Lai}
\email{yufangyuan@sjtu.edu.cn, dl57@cornell.edu}

\author[0009-0004-6973-3955]{Fangyuan Yu}
\affiliation{Zhiyuan College, Shanghai Jiao Tong University, Shanghai 200240, China}
\affiliation{Tsung-Dao Lee Institute, Shanghai Jiao Tong University, Shanghai 201210, China}

\author[0000-0002-1934-6250]{Dong Lai}
\affiliation{Center for Astrophysics and Planetary Science, Department of Astronomy, Cornell University, Ithaca, NY 14853, USA}
\affiliation{Tsung-Dao Lee Institute, Shanghai Jiao Tong University, Shanghai 201210, China}

\begin{abstract}
In galactic centers, stars and binaries can be injected into low-angular-momentum orbits, resulting in close encounters with the central supermassive black hole (SMBH). 
Previous works have shown that under different conditions, such close encounters can lead to the break-up of the binary, disruptions of both stars and collision between the stars. 
We use 3-body scattering experiments to characterize these different outcomes for a range of system parameters, such as $\beta_b$, the ratio of binary tidal radius to pericenter distance $r_p$ to the SMBH and the compactness of the binary.
We focus on stellar collisions, which occur for a range of $\beta_b$'s, with a few to 10's percent probabilities (depending on the compactness of the binary). 
In gentle encounters ($\beta_b\lesssim 1$), stellar collisions occur after the pericenter passage, and the merger remnants are typically ejected from the SMBH at a small velocity. 
In deep encounters ($\beta_b\gtrsim 1$), collisions occur near the pericenter, with the impact velocity a few times the escape velocity of the star, and the merger remnants are typically bound to the SMBH. 
We suggest that stellar collisions induced by binary-SMBH encounters may produce exotic stars in galactic centers, trigger accretion flares onto the SMBH due to the mass loss, and result in bound merger remnants causing repeated partial TDEs.
\end{abstract}

\keywords{Black holes; Binary stars; Tidal disruption; Stellar dynamics}

\section{Introduction} 
\label{sec:intro}

The unique conditions at the centers of galaxies, namely a large number of high-velocity stars in the deep gravitational potential well of a central supermassive black hole (SMBH), can lead to unusual patterns of star formation, dynamical evolution and nuclear activities \citep[e.g.,][]{Alexander2017ARAA}.
Within the sphere of influence of the SMBH, two-body relaxation and other collective processes are efficient, and stars in the nuclear cluster can be ``kicked" into low-angular-momentum orbits toward the SMBH, leading to tidal disruptions of the stars \citep{Rees1988nature,Stone2020SSRv}. 
Since the 1990s, many tidal disruption events (TDEs) have been detected \citep{Komossa2015JHEA,Van2020SSRv,Gezari2021ARAA}.
Similar to TDEs, a stellar binary that passes sufficiently close to the SMBH is tidally disrupted, resulting in the ejection of a binary component with very high velocity \citep{Hills1988Nature}. 
Many hyper-velocity stars (HVSs) flying away from the Galactic Center have been found \citep[e.g.][]{Brown2005ApJ,Koposov2019MNRAS}.
The velocities of these HVSs are so high (measured at $500-1000$ km/s) that binary break-ups by the Galactic Center SMBH provide a natural explanation \citep{Yu2003ApJ,Gould2003ApJ}.
A binary break-up also leaves behind a star on a bound, eccentric orbit around the SMBH. 
It has been suggested that such a star may be related to the phenomena of partial TDEs \citep{cufari2022ApJL,Somalwar2023arXiv} 
and quasi-periodic eruptions \citep{Lu2023MNRAS,Linial2023ApJ,Zhou2024PRD}.

Many previous works have studied the dynamics of close encounters between stellar binaries and SMBHs \citep[e.g.,][]{Gualandris2005MNRAS,Bromley2006ApJ,Ginsburg2006MNRAS,Sesana2007ApJ,Sari2010ApJ,Brown2018MNRAS}. 
In addition to the binary break-up that produces an HVS and a tightly bound star, other outcomes are possible. 
When the pericenter distance of the binary to the SMBH is comparable to the tidal radius of the star, 
both members of the binary will be disrupted sequentially; such double TDEs may exhibit light curves that are distinct from single TDEs \citep{Mandel2015ApJL,Bradnick2017MNRAS}.
During or shortly after the binary-SMBH close encounter, the two stars may come so close to each other that a physical collision occurs \citep{Ginsburg2007MNRAS,Antonini2010ApJ,Bradnick2017MNRAS}; 
such a collision leads to stellar merger and mass loss, which may produce exotic stars and generate observable electromagnetic signatures.

This paper is devoted to a systematic study of the various possible outcomes of the close encounters between stellar binaries and SMBHs,
including binary break-up, double stellar disruptions and stellar collision/merger. 
We extend previous works by surveying a wide parameter space, 
for encounter pericenter distances ranging from outside the binary's tidal sphere to deep inside it.
Using restricted 3-body scattering experiments, we characterize the orbital energy distribution of the stars after the binary break-up, and determine the time separation between the two stellar disruptions in double TDEs. 
For stellar collisions, we are interested in questions such as
(a) what is the probability of stellar collisions for different system parameters?
(b) what is the velocity distribution of the collisions?
(c) how likely are the merger remnants bound to the SMBH, and what are their orbital energies?
Overall, an important goal of our study is to obtain the key results related to binary-SMBH encounters that can be scaled to different system parameters. 

Our paper is organized as follows. 
In Section~\ref{sec:method and setup} we describe the key parameters of the problem and the setup of our numerical experiments. 
In Section~\ref{sec:binary tidal break-up} we revisit the Hills mechanism of binary break-ups. 
In Section~\ref{sec:double stellar disruption} we present numerical results on double stellar disruptions.
In Section~\ref{sec:stellar merger} we present the results for stellar collisions/mergers and merger remnants.
We summarize our key findings and discuss some potential applications in Section~\ref{sec:application and summary}.

\section{Method and Setup}
\label{sec:method and setup}

We use the $N$-body code \texttt{REBOUND} with the \texttt{IAS15} integrator \citep{rebound,reboundias15} to simulate the close encounter between a binary and an SMBH.

\subsection{Key Parameters and Scalings}
\label{sec:key parameters}

We introduce several key parameters and scalings in the binary-SMBH system to simplify the problem. 

A binary is tidally disrupted when the tidal force of the SMBH dominates over the self-gravity of the binary. 
The characteristic distance at which this occurs is the binary tidal radius:
\begin{equation}
r_{\rm tide}^{b} \equiv a_b \Bigl(\frac{M}{m_{12}} \Bigr)^{1/3},
\label{eq:r_tide^b}
\end{equation}
where $a_b$ is the semi-major axis of the binary's initial orbit, $M$ is the mass of the SMBH and $m_{12}\equiv m_1+m_2$ is the total mass of the binary.
A star is tidally disrupted when its distance to the SMBH is less than the stellar tidal radius:
\begin{equation}
r_{\rm tide}^{\star} \equiv R_\star \Bigl(\frac{M}{m} \Bigr)^{1/3},
\label{eq:r_tide^s}
\end{equation}
where $R_\star$ is the radius of the star and $m$ is the mass of the star.
For tight binaries, there is a non-negligible chance that close encounters with SMBH can lead to stellar collisions.
We denote $R_{\rm col}$ as the distance between the two stars when they collide. 
For a binary with identical stars, $R_{\rm col} = 2R_\star$.
In our numerical experiments, we assume the binary approaches the SMBH on a near-parabolic orbit with pericenter distance $r_p$.

From the length scales mentioned above, we define three dimensionless parameters:
\begin{equation}
    \alpha \equiv a_b/R_{\rm col},
    \label{eq:alpha}
\end{equation}
which describes the ``compactness" of the binary,
\begin{equation}
    \beta_b \equiv r_{\rm tide}^b/r_p,
    \label{eq:beta}
\end{equation}
which describes the ``depth" of the encounter from the binary perspective, and
\begin{equation}
    \beta_\star \equiv r_{\rm tide}^\star/r_p,
    \label{eq:beta_s}
\end{equation}
which specifies the ``depth" of the encounter from the star's perspective.
(For a binary with identical stars, $\beta_b=2^{2/3}\alpha\beta_\star$.)
The dynamics and outcomes of the binary-SMBH encounter then depend on the three dimensionless parameters
(as well as several angles that specify the geometry and phase of the encounter; see below).
In our study, we consider $\alpha \gtrsim$ a few, so that $\beta_b > $ (a few) $\times \beta_\star$.

Note that for $r_p \gg a_b$, or equivalently $(M/m_{12})^{1/3} \gg \beta_b$, the mass ratios $M/m_{12}$ and $m_2/m_1$ do not directly affect the dynamics of the encounter.
To see this, consider the equation of motion:
\begin{equation}
\ddot{\boldsymbol r}_{21}=-\frac{GM}{r_2^3}{\boldsymbol r_2}+\frac{GM}{r_1^3}{\boldsymbol r_1}-\frac{Gm_{12}}{r_{21}^3}{\boldsymbol r_{21}},
\label{eq:EoM21}
\end{equation}
where ${\boldsymbol r_1}$ and ${\boldsymbol r_2}$ are the respective distance vectors of $m_1$ and $m_2$ from the SMBH $M$,
${\boldsymbol r_{21}\equiv {\boldsymbol r_2}-{\boldsymbol r_1}}$ is the relative distance vector between the stars. 
For $r_{21} \ll r_1$ and $r_2$, we can expand ${\boldsymbol r_1}$ and ${\boldsymbol r_2}$ around ${\boldsymbol r_{\rm CM}}$, which is the binary center-of-mass position vector (relative to the SMBH). 
This gives:
\begin{equation}
\ddot{\boldsymbol r}_{21} \simeq -\frac{GM}{r_{\rm CM}^3}{\boldsymbol r_{21}}+3\frac{GM}{r_{\rm CM}^5}({\boldsymbol r_{21}}\cdot{\boldsymbol r_{\rm CM}}){\boldsymbol r_{\rm CM}}-\frac{Gm_{12}}{r_{21}^3}{\boldsymbol r_{21}}.
\label{eq:EoMlinear}
\end{equation}
We can further rescale Equation~(\ref{eq:EoMlinear}) by defining the length unit $r_0\equiv(m_{12}/M)^{1/3}r_p$ and the time unit $t_0\equiv\sqrt{r_p^3/(GM)}$.
Let $\tilde{\boldsymbol r} \equiv {\boldsymbol r_{21}}/r_0$, ${\tilde t} \equiv t/t_0$, and $\hat{\boldsymbol r}_{\rm CM}={\boldsymbol r}_{\rm CM}/r_{\rm CM}$, we have \citep[cf.][]{Sari2010ApJ}
\begin{equation}
\frac{{\rm d^2} \tilde{\boldsymbol r}}{{\rm d} \tilde{t}^2}=\Bigl(\frac{r_p}{r_{\rm CM}}\Bigr)^3[-\tilde{\boldsymbol r}+3(\tilde{\boldsymbol r}\cdot\hat{\boldsymbol r}_{\rm CM})\hat{\boldsymbol r}_{\rm CM}]-\frac{\tilde{\boldsymbol r}}{\tilde{r}^3}.
\label{eq:EoMscaling}
\end{equation}
For parabolic encounters, $r_p/r_{\rm CM}$ and $\hat{\boldsymbol r}_{\rm CM}$ depend explicitly on $\tilde{t}$, i.e.,
\begin{equation}
    \frac{r_p}{r_{\rm CM}}=\frac{1+\cos f}{2},
\label{eq:rCM}
\end{equation}
\begin{equation}
    \tilde{t}=\frac{\sqrt{2}}{3} \tan \left ( \frac{f}{2} \right ) \Bigl[ 3+\tan^2 \left (\frac{f}{2} \right ) \Bigr],
\label{eq:ttof}
\end{equation}
where $f$ is the true anomaly.

From Equation~(\ref{eq:EoMscaling}), we clearly see that the dynamics of binary-SMBH encounters does not depend on the mass ratios $M/m_{12}$ and $m_2/m_1$ as long as the condition $r_{21} \ll r_{\rm CM}\simeq r_1 \simeq r_2$ is satisfied.
Since the outcome of the encounter is mainly determined around $r_{\rm CM} \sim r_p$, it does not depend on $M/m_{12}$ and $m_2/m_1$ explicitly.
Of course, in the case when $m_1$ and $m_2$ become separated after the binary-SMBH encounter, the post-encounter evolution of $m_1$ and $m_2$ may not be described by Equation~(\ref{eq:EoMscaling}).

Using Equation~(\ref{eq:EoMscaling}) to study the outcomes of binary-SMBH encounters requires three dimensionless parameters: 
$a_b/r_0=\beta_b$ is needed for specifying the initial conditions, the collision distance $R_{\rm col}/r_0=\alpha^{-1}\beta_b$ is needed for considering stellar collisions, and $R_\star/r_0=\beta_\star$ is needed for considering stellar tidal disruptions. 
In our study, we will survey the outcomes of binary-SMBH encounters in the $(\alpha, \beta_b, \beta_\star)$ parameter space.

Note that in this paper we are considering the first close encounter between a SMBH and a relatively tight binary, with $\alpha\lesssim 15$. Thus, unless otherwise noted, we will assume that the initial binary orbit is circular. This is reasonable because tidal interaction between the binary components is expected to have circularized the orbit. Indeed, 
\citet{Meibom2005ApJ} found that the characteristic circularization period of solar-type MS stars is $8-15$ days (depending on the stellar ages), which corresponds to $\alpha =11-16$.

\subsection{Simulation Setup}
\label{sec:setup}

We consider binary stars $m_1$ and $m_2$, initially on a circular orbit with semi-major axis $a_b$. 
The center of mass of the binary approaches the SMBH $M$ on a nearly parabolic orbit.
We already know from Equation~(\ref{eq:EoMscaling}) that $M/m_{12}$ and $m_2/m_1$ do not affect the dynamics of the encounter as long as $M/m_{12}\gg 1$.
Thus throughout the paper, unless otherwise noted, our simulations adopt $m_1=m_2=m$ and $M/m = 10^6$. 

In addition to the pericenter distance $r_p$ (or equivalently $\beta_b$), the trajectory of the binary relative to the SMBH is specified by the velocity at infinity $v_\infty$.
Note that we are considering ``hard" binaries so that $v_\infty$ is much less than the binary orbital velocity $v_{\rm orb}\equiv\sqrt{Gm_{12}/a_b}$.
In our fiducial simulations, we adopt $v_\infty/v_{\rm orb}=0.1$ for all encounters. 
We have tested that a smaller value of $v_\infty/v_{\rm orb}$ does not affect our results.

In addition to the three dimensionless parameters $\alpha \equiv a_b/R_{\rm col}$, $\beta_b \equiv r_{\rm tide}^b/r_p$ and $\beta_\star \equiv r_{\rm tide}^\star/r_p$, the outcome of an encounter depends on various angles, including the inclination angle ($i$) between the binary-SMBH orbital plane and the binary plane, the argument of periastron ($\omega$) of the binary orbit, and the initial orbital phase ($\lambda$) of the binary. 
Note that since the initial binary orbit is circular, the outcome of an encounter does not depend on $\Omega$, the longitude of the node of the binary orbit. 
In our simulations, we sample the dimensionless parameter $\beta_b$ ranging from $0.5$ to $10$ (for $\beta_b \lesssim 0.4$, almost all binaries survive). 
For the three angles, we sample
\begin{enumerate}
    \item $i$, with a $\sin (i)$ probability distribution, between 0 and $\pi$, or equivalently, uniform distribution in $\cos (i)$,
    \item $\omega$, with a flat prior between 0 and $2\pi$,
    \item $\lambda$, with a flat prior between 0 and $2\pi$.
\end{enumerate}
Each simulation covers a time span from 20\,$t_b$ before the pericenter passage of the SMBH (approximately 15\,$r_{\rm tide}^b$ away) to 20\,$t_b$ after, where $t_b$ is defined as $t_b \equiv \sqrt{a_b^3/(Gm_{12})}$. 

While most of this paper deals with initially circular binaries, for completeness, we will also present some results for binaries with initially finite eccentricity,  $e_b$.
When considering elliptical binary orbits, the outcome of an encounter additionally depends on the longitude of the ascending node of the binary orbit ($\Omega$). In our simulations, we sample 
$\Omega$ with a flat prior between 0 and $2\pi$.
Furthermore, because the phase variation of an elliptical orbit is not uniform, we no longer employ a flat prior when sampling the initial orbital phase ($\lambda$). 
Instead, we use the orbital phase values derived from a flat prior in the orbital time over the interval from  $0$ to $P_b$, the binary orbital period.

\subsection{Outcomes and Branching Criteria}
\label{sec:outcomes and branching criteria}

After a close binary-SMBH encounter, there are four possible outcomes if the finite size of the star is taken into account.
It is important to establish appropriate criteria for distinguishing the different outcomes: 

\begin{itemize}
    \item {\it Surviving binary}: 
    When $r_p \gtrsim 2 r_{\rm tide}^b$, i.e., $\beta_b \lesssim 0.5$, the binary is not disrupted by the SMBH, although its orbit may be significantly disturbed.
    For a surviving binary, the separation between the two stars $r_{12}$ is always greater than the collision distance $R_{\rm col}$ during the whole encounter, 
    and the minimum distance between each star and the SMBH is larger than $r_{\rm tide}^\star$.
    At the end of the simulation, $r_{12}$ is less than the ``instantaneous" Hill radius $r_{\rm Hill}=(m_{\rm 12}/3M)^{1/3}\,d$, where $d$ is the minimum separations between each star and the SMBH; 
    the relative energy between the two stars is negative, i.e.,
    \begin{equation}
        \frac{1}{2}{\bf v}_{12}^2-\frac{Gm_{12}}{r_{12}}<0,
        \label{Eq:inertial}
    \end{equation}
    where ${\bf v}_{12}$ is the relative velocity between the two stars.

    \item {\it Binary tidal break-up}: 
    When $r_p \lesssim r_{\rm tide}^b$, i.e., $\beta_b \gtrsim 1$, the binary is broken up by the tidal field of the SMBH.
    The potential energy difference across the binary results in one member being ejected, becoming an HVS, while the SMBH captures the other member into an eccentric orbit \citep{Hills1988Nature}.
    In this case, $r_{12}$ is always greater than $R_{\rm col}$ and the minimum distance between the star and SMBH is larger than $r_{\rm tide}^\star$.
    We calculate the energy of star $m_i$ relative to an SMBH, via
    \begin{equation}
        \varepsilon_i\equiv \frac{1}{2} ({\bf v}_i-{\bf V})^2-\frac{GM}{r_i},
        \label{Eq:energy}
    \end{equation}
    where ${\bf v}_i$ is the star's velocity and ${\bf V}$ is the velocity of the SMBH.
    One of the stars has an energy $\varepsilon_i>0\,$ (ejected as an HVS), and the other has $\varepsilon_i<0$ (captured by the SMBH).
    
    \item {\it Stellar disruption}:
    When $r_p \lesssim r_{\rm tide}^\star$ or $\beta_\star \gtrsim 1$, the binary first undergoes tidal break-up at $r_1\simeq r_2\simeq r_{\rm CM} \sim r_{\rm tide}^b$, becoming two nearly independent stars. 
    These stars then move to smaller distances and are tidally disrupted when $r_1, r_2\simeq r_{\rm tide}^\star$.
    In this case, two separate tidal disruption events will occur in sequence \citep{Mandel2015ApJL}.
    
    \item {\it Stellar collision}: 
    In the aforementioned parameter space, if the finite size of the star is taken into account, the collision of the two stars becomes possible. 
    For $\beta_b \sim 1$, if the orbit of the surviving binary has a large eccentricity, the stars may collide after the pericenter passage; 
    for $\beta_b \gg 1$, a stellar collision may occur dynamically during the binary's pericenter passage very close to the SMBH.
    In this case, the minimum distance between each star and SMBH is larger than $r_{\rm tide}^\star$,
    and the minimum distance between the two stars $r_{12}$ during the whole encounter is less than $R_{\rm col}$.
\end{itemize}

We note that under the above classification scheme of the encounter outcomes, stellar collisions and stellar TDEs may occur simultaneously in some cases. 
However, the situation becomes very complex once either collision or TDE occurs, 
and hydrodynamic simulations are required to fully understand these interactions. 
Therefore, in this paper, we will mostly focus on the range of parameters that avoid stellar TDEs while considering stellar collisions, i.e., we will consider $r_p\gtrsim r_{\rm tide}^\star$.
For binaries with identical stars, this implies 
\begin{equation}
    \beta_b \lesssim 2^{2/3} \alpha, \qquad (r_p \gtrsim r_{\rm tide}^\star).
\label{eq:avoidTDE}
\end{equation}

\section{Binary Tidal Break-Up}
\label{sec:binary tidal break-up}

The Hills mechanism for binary break-up is well understood \citep{Hills1988Nature}. 
In this section, for completeness, we analyze the binary break-up fraction and the energy distribution of the stars produced by this mechanism.

\begin{figure}[htbp]
\centering
\includegraphics[width=1.02\columnwidth]{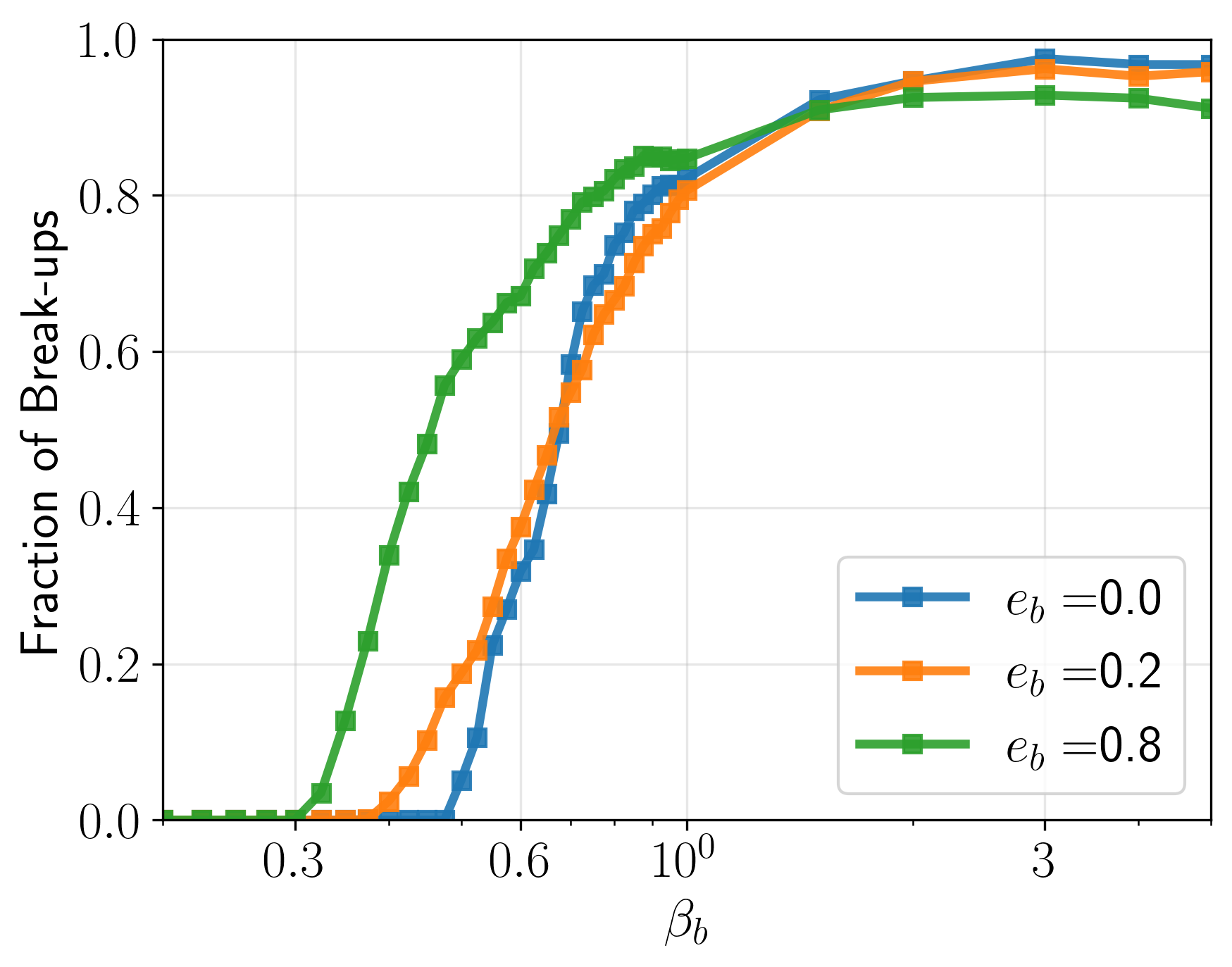}
\caption{
The fraction of binary break-ups as a function of the dimensionless parameter $\beta_b\equiv r_{\rm tide}^b/r_p$ for several different values of the initial binary eccentricity.}
\label{Fig:Diffe_Break_up_frac}
\end{figure}

Figure~\ref{Fig:Diffe_Break_up_frac} shows the fraction of binary break-ups following an encounter between the binary and the SMBH.
We see that binaries on (initially) elliptical orbits are more susceptible to tidal break-up. 
Moreover, even when the pericenter distance is smaller than the binary tidal radius ($\beta_b > 1$), the break-up fraction does not reach unity \citep[see also][]{Sari2010ApJ}.

Consider the energy (relative to the SMBH) of each star after the binary break-up.
At $r_{\rm CM}=r_{\rm tide}^b$, the distance of $m_1$ to the SMBH has a spread of $2a_{b1}=2(m_2/m_{12})a_b$.
The characteristic tidal energy is
\begin{equation} 
E_{\rm tide1} \equiv \frac{GM m_1}{({r_{\rm tide}^{b}})^2} (2a_{b1})=\frac{2GM \mu_{12}a_b}{({r_{\rm tide}^{b}})^2},
\label{eq:Etide}
\end{equation}
where $\mu_{12}=m_1m_2/m_{12}$. Note that $E_{\rm tide1}=E_{\rm tide2}=E_{\rm tide}$, so we only need to consider $m_1$.

\begin{figure}[htbp]
\centering
\includegraphics[width=1.02\columnwidth]{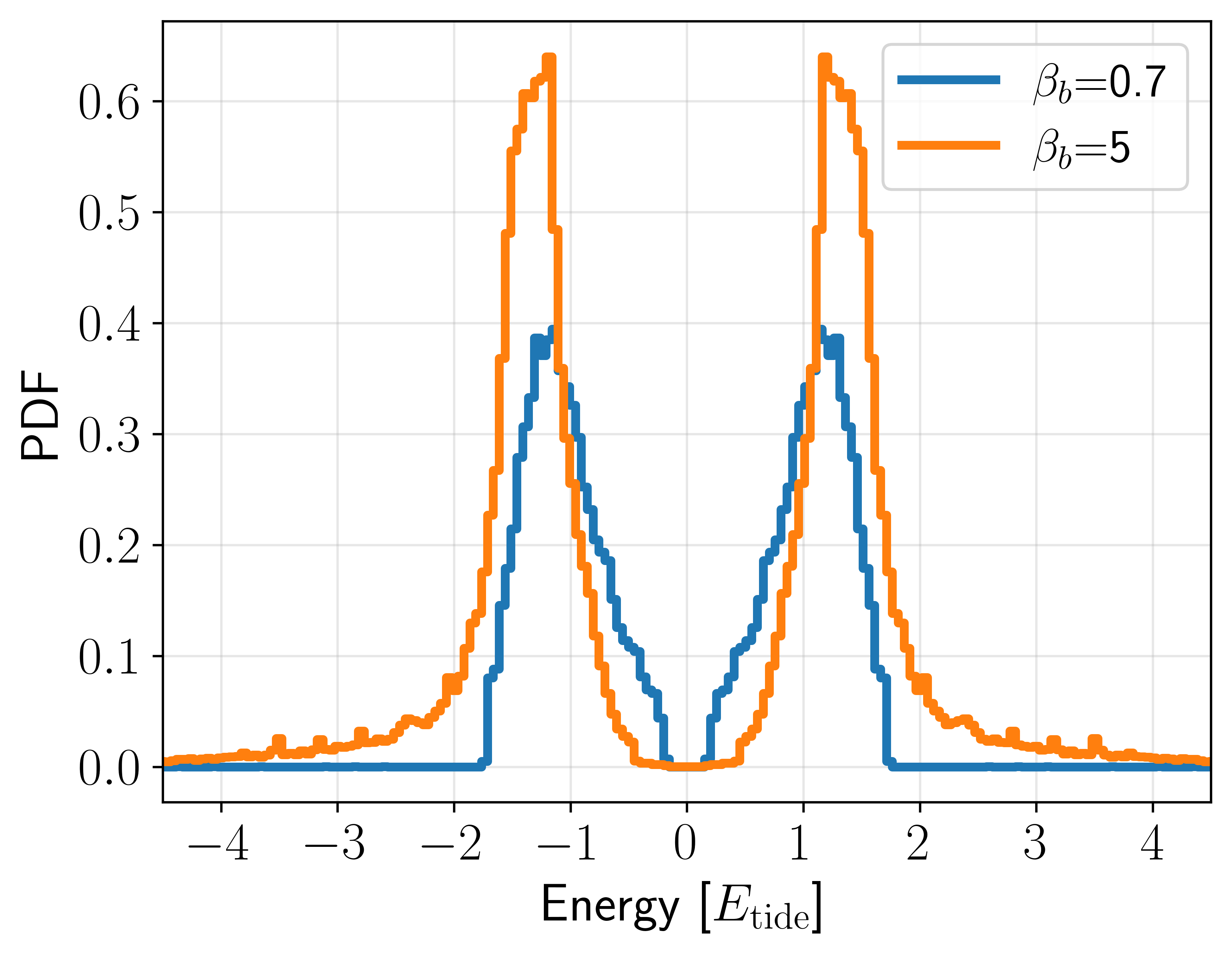}
\caption{
Probability density function (PDF) of the orbital energy of each star (in units of $E_{\rm tide}=E_{\rm tide1}$, given by Equation~\ref{eq:Etide}) after binary break-up. The initial binary has $e_b=0$.
The blue and orange solid lines represent the results for two different values of $\beta_b$: gentle encounters ($\beta_b=0.7$) and deep encounters ($\beta_b=5$). 
These PDFs are obtained by sampling all encounters with different phase angles and inclinations.
Note that the PDF does not normalized to unity, because for a given $\beta_b$, only a fraction of the encounters lead to binary break-ups; 
each PDF includes all encounters, not just those leading to binary break-ups.}
\label{Fig:Hills_Edistribution}
\end{figure}

Figure~\ref{Fig:Hills_Edistribution} shows the probability density function of $E_1$ (the energy of $m_1$ relative to the SMBH) after the tidal break-up for two $\beta_b$ values, $\beta_b=0.7$ and $5$. 
The former represents gentle encounters, with only a small fraction of binaries being tidally disrupted, while the latter represents deep encounters, with most binaries being disrupted.

From the figure, we can see that the energy distribution exhibits clear bimodality. 
The peaks are at $|E_{1,2}|\simeq 1.3 E_{\rm tide}$, and depend weakly on $\beta_b$.
In the case of $\beta_b=0.7$, there is a significant truncation of energy at $|E_{1,2}|\simeq 1.6E_{\rm tide}$. 
In contrast, for $\beta_b=5$, there is a noticeable long tail, indicating that deep encounters are more likely to produce HVSs with extreme velocities.

\begin{figure}[htbp]
\centering
\includegraphics[width=1.02\columnwidth]{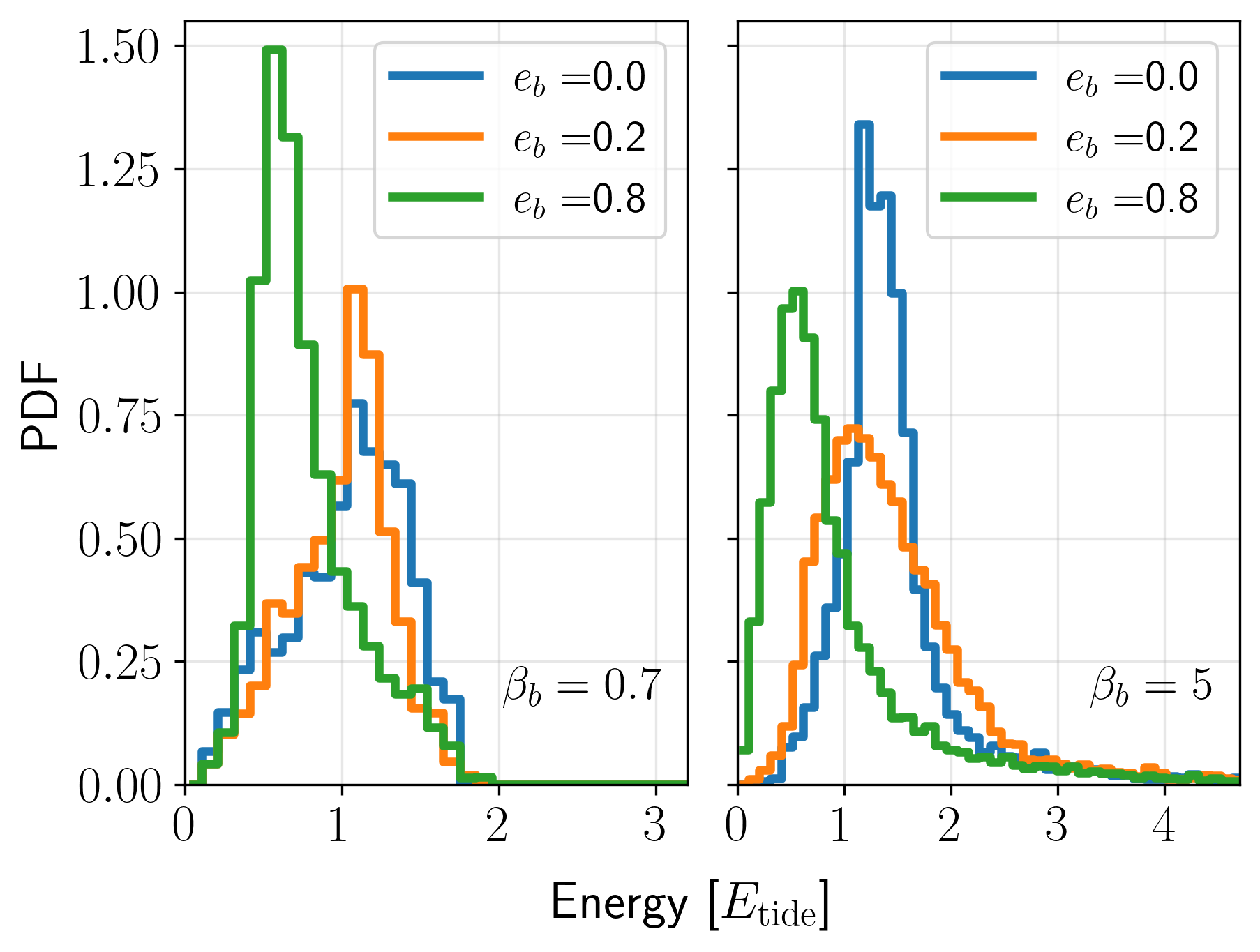}
\caption{
Same as Figure~\ref{Fig:Hills_Edistribution}, except showing only the positive part of the energy (the negative part is perfectly symmetric with respect to the positive part) and for different initial binary eccentricities. 
The left panel represents gentle encounters ($\beta_b=0.7$), while the right panel represent deep encounters ($\beta_b=5$).}
\label{Fig:Diffe_Hills_E}
\end{figure}

Figure~\ref{Fig:Diffe_Hills_E} shows the distributions of $E_1$ for binaries with different initial eccentricities.  
We see that for small (but finite) eccentricities, the peak of the distribution slightly decreases (compared to the $e_b=0$ case), while the distribution itself becomes broader. 
This can be explained by changing $(2a_{b1})$ in Eq.~(\ref{eq:Etide}) to the range between $[2a_{b1}(1-e_b)]$ and $[2a_{b1}(1+e_b)]$.
For larger eccentricities, the peak of the energy distribution becomes significantly lower.

It is of interest to compare binary break-up and stellar tidal disruption. 
In the latter case, the disrupted stellar debris has specific energy $\varepsilon$ ranging from $-\varepsilon_{\rm tide}$ to $\varepsilon_{\rm tide}$, with $\varepsilon_{\rm tide} \simeq GMR_\star/(r_{\rm tide}^\star)^2$. 
The distribution of the debris energy (mass per unit $\varepsilon$) ${\rm d} m/{\rm d} \varepsilon$ is often assumed to be a top-hat \citep{Rees1988nature}. 
In reality, ${\rm d} m/{\rm d} \varepsilon$ depends on the density profile of the star, and more weakly on $\beta_\star$ \citep[e.g.,][]{Lodato2009MNRAS,Guillochon2013ApJ};
${\rm d} m/{\rm d} \varepsilon$ becomes a constant for $|\varepsilon| \ll \varepsilon_{\rm tide}$ and peaks at $\varepsilon = 0$. 
In the case of a binary tidal break-up, the PDF of disrupted stellar energy is qualitatively different, as there is a gap at $|E_{1,2}|\sim 0$, or $|E_{1,2}| \ll |E_{\rm tide}|$.
This difference is expected: there is no material in the space ``between the stars", and for a given $\beta_b$, there is always a chance that the binary is not disrupted --- such surviving binary will have rather small ($\ll E_{\rm tide}$) orbital energy around the SMBH after the close encounter.

\section{Double stellar disruptions}
\label{sec:double stellar disruption}

In very deep encounters, where $r_p$ is much less than $r_{\rm tide}^b$ ($\beta_b \gg 1$) and becomes comparable to $r_{\rm tide}^\star$ ($\beta_\star \sim 1$), the interaction typically results in the sequential disruption of the two stars. 
This possibility has been discussed by \citet{Mandel2015ApJL}. 
Here our main concern is the time separation $\Delta t_{\rm DD}$ between the double disruptions (DD) and whether we can distinguish between the two stellar disruption flares if double stellar disruptions occur.

A simple estimate of $\Delta t_{\rm DD}$ goes as follows. 
When a binary reaches the distance $r_{\rm tide}^b$ from the SMBH, the tidal force of the SMBH starts to dominate over the mutual stellar interaction. 
As a result, the individual star begins to approach the SMBH independently along nearly identical paths. 
A tidal disruption event occurs when the star reaches $r_{\rm tide}^\star$ from the SMBH. 
The journey from $r_{\rm tide}^b$ to $r_{\rm tide}^\star$ takes about the same amount of time for each star. 
Therefore, we only need to determine the time difference between the two stars as each reaches $r_{\rm tide}^b$.
At $r_{\rm CM} \sim r_{\rm tide}^b$, the maximum tangential stellar separation is $a_b$, and the star's velocity is $v_{\rm CM} \sim \sqrt{GM/r_{\rm tide}^b}$ [Recall that $v_{\rm CM}$ is much larger than the internal orbital velocity of the binary: $v_{\rm CM}/v_{\rm orb}=(M/m_{12})^{1/3} \gg 1$]. 
Thus, the maximum time difference $(\Delta t_{\rm DD})_{\rm max}$ for double disruption is:
\begin{equation}
(\Delta t_{\rm DD})_{\rm max} \sim \frac{a_b}{\sqrt{GM/r_{\rm tide}^b}}\sim \Bigl(\frac{m_{12}}{M}\Bigr)^{1/3}\sqrt{\frac{a_b^3}{Gm_{12}}}.
\label{eq:deltaT}
\end{equation}

\begin{figure}[htbp]
\centering
\includegraphics[width=1.02\columnwidth]{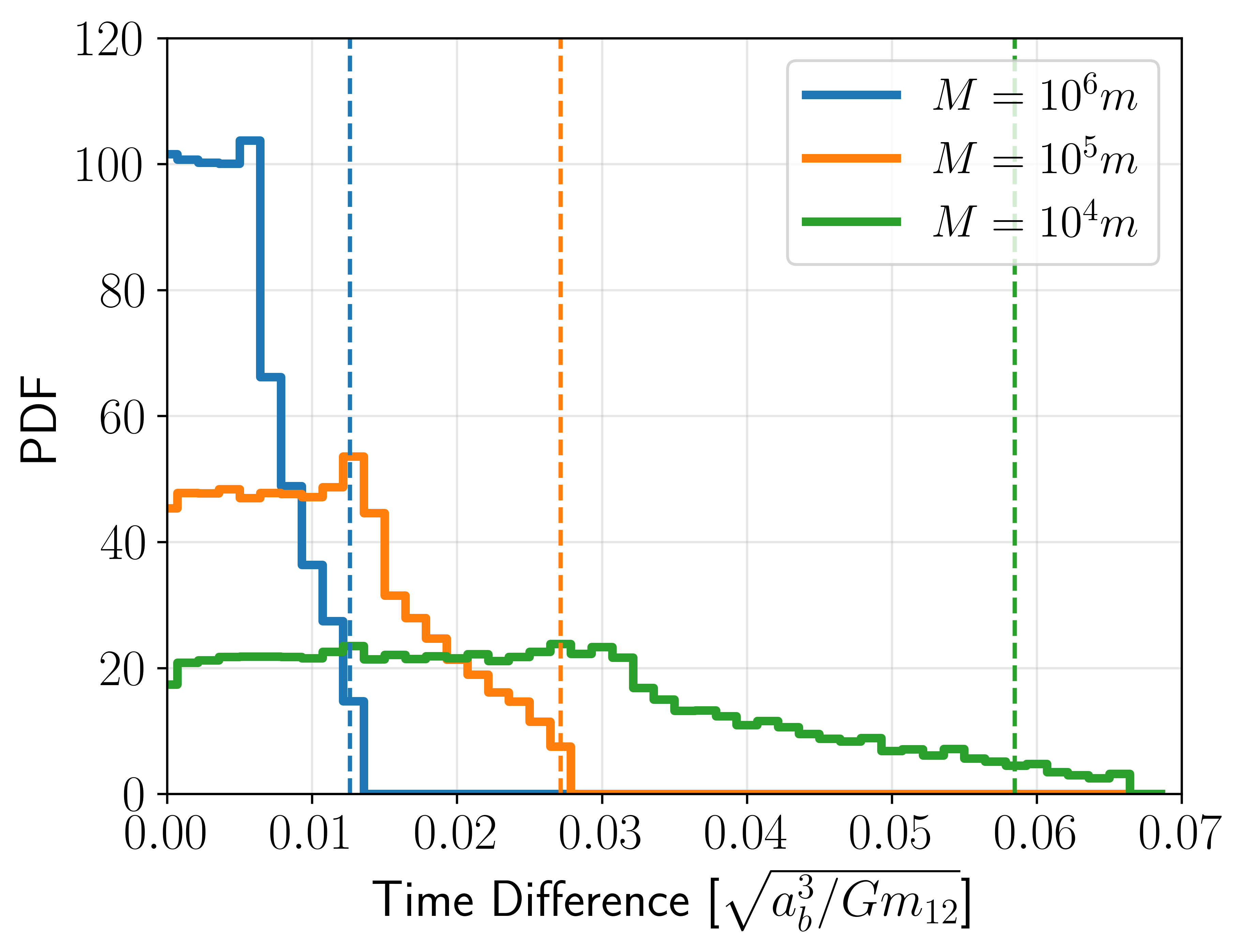}
\caption{
Probability density function of the time difference $\Delta t_{\rm DD}$ between double stellar disruptions for $\alpha = 8, \beta_b = 20$.
The blue, orange, and green solid lines represent the results for $M/m=10^4, 10^5$ and $10^6$ (assuming $m_1=m_2=m$), respectively, while the vertical dashed lines represent $(\Delta t_{\rm DD})_{\rm max}$ given by Equation~(\ref{eq:deltaT}).}
\label{Fig:deltaT}
\end{figure}

Figure \ref{Fig:deltaT} shows our numerical results (assuming $m_1=m_2=m$) for the distribution of $\Delta t_{\rm DD}$ for encounters with $\alpha = 8, \beta_b = 20$ (Note that these parameters violate the condition in Equation~\ref{eq:avoidTDE}).
To test the mass scaling, we consider $M/m=10^4, 10^5$ and $10^6$.
We see that there is indeed an upper bound on $\Delta t_{\rm DD}$ in the simulation data and that the upper bound agrees with Equation~(\ref{eq:deltaT}). 
The distribution of $\Delta t_{\rm DD}$ is nearly uniform between $0$ and $(\Delta t_{\rm DD})_{\rm max}/2$ and decreases to zero towards $(\Delta t_{\rm DD})_{\rm max}$. 
Importantly, $\Delta t_{\rm DD}$ is much smaller than the orbital period of the binary (see Eq.~\ref{eq:deltaT}). 
Since the occurrence of double stellar disruptions requires $\beta_b \gtrsim \alpha$, it tends to occur primarily for tight binaries (small $\alpha$) with shorter orbital periods.

Note that although $\Delta t_{\rm DD}$ is small, if the stars have different masses/radii, their disrupted bound streams will return to the pericenter on different timescales \citep{Mandel2015ApJL}. 
Indeed, the return time of the most bound debris is 
\begin{equation}
    t_{\rm min}
\sim {\pi M\over m} \left(\frac{R_\star^3}{2GM}\right)^{1/2}.
\end{equation}
For $R_\star\propto m^{0.8}$, we have $t_{\rm min} \propto m^{0.2}$, which means that for binaries with unequal masses, we may see a double-peaked burst.
In addition, the interaction between the two stellar streams may give rise to dissipation and electromagnetic radiation signals \citep{Bonnerot2019MNRAS}.

\section{Stellar collision}
\label{sec:stellar merger}

We now examine stellar collisions resulting from close encounters between binaries and SMBHs.
Previous work has shown that collisions can happen during deep ($\beta_b \gg 1$) encounters \citep{Ginsburg2007MNRAS}. 
For a binary moving around the SMBH over many orbits, the Lidov-Kozai effect can induce very high eccentricity in the binary orbit and thus stellar collisions \citep{Antonini2010ApJ,Bradnick2017MNRAS}. 
Here we restrict to stellar collisions induced by a single (``first'') close encounter between the circular binary and the SMBH.

Due to the excitation of eccentricity during close encounters, it is likely that a binary will have a finite eccentricity when it approaches the black hole multiple times, increasing the probability of stellar collisions \citep{Antonini2010ApJ,Bradnick2017MNRAS}.

In Section~\ref{sec:collision fraction} we determine the parameter space where the collision occurs and the corresponding fraction of collisions;
in Section~\ref{sec:collision velocity} we analyze the collision velocities; 
in Section~\ref{sec:properties of merger remnant} we examine the properties of the merger remnants.

\subsection{Collision Fraction and Parameter Space}
\label{sec:collision fraction}

To analyze the fraction of stellar collisions, we divide the $\beta_b$ parameter space into two regimes:
gentle encounters ($\beta_b \lesssim 1$) where most binaries are not disrupted, and deep encounters ($\beta_b \gtrsim 1$) where most binaries are disrupted.

\subsubsection{Gentle Encounters (\texorpdfstring{$\beta_b \lesssim 1$}{β<1})}
\label{sec:br_regime1}

For gentle encounters, most binaries survive the pericenter passage. 
However, some systems can achieve high eccentricities, causing the binary stars to collide shortly after the pericenter passage. 

\begin{figure}[htbp]
\centering
\includegraphics[width=1.02\columnwidth]{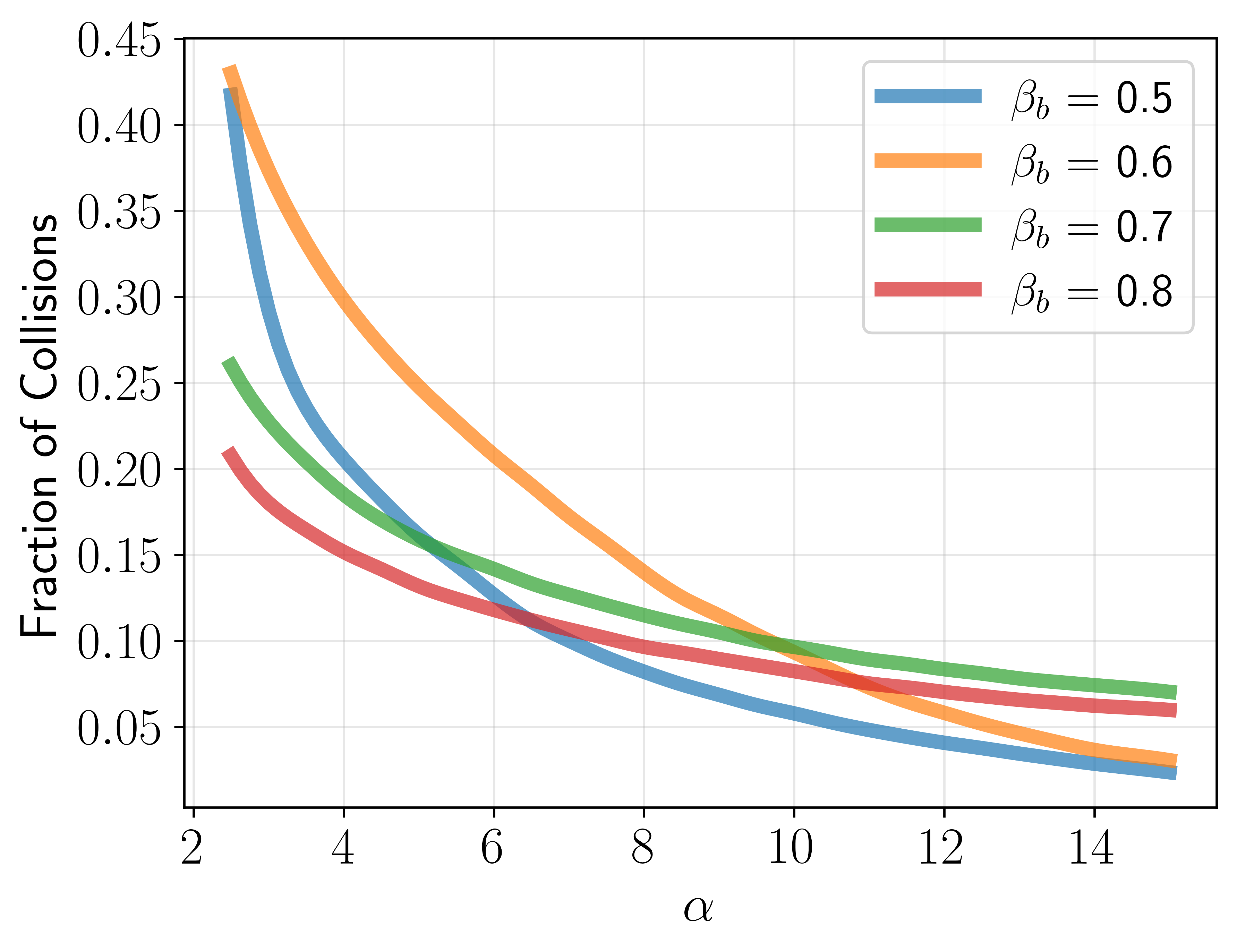}
\caption{
The fraction of stellar collisions induced by binary-SMBH encounters as a function of the dimensionless parameter $\alpha \equiv a_b / R_{\rm col}$, for several different values of $\beta_b \equiv r_{\rm tide}^b/r_p=0.5, 0.6, 0.7$ and $0.8$ (all are gentle encounters).}
\label{Fig:col_frac_lowbeta}
\end{figure}

Figure~\ref{Fig:col_frac_lowbeta} shows the fraction of stellar collisions as a function of $\alpha \equiv a_b / R_{\rm col}$, for simulations with $\beta_b=0.5, 0.6, 0.7$ and $0.8$.
We see that for a given $\beta_b$, the collision fraction decreases monotonically with increasing $\alpha$, as expected.
However, for a given $\alpha$, the collision fraction initially rises and then falls as $\beta_b$ increases from $0.5$ to $0.8$. 
This is due to a combination of two factors: the orbital eccentricity that the binary can attain after the pericenter passage (see below) and the proportion of binaries that survive the close encounter.
When $\beta_b$ is small (e.g., $\beta_b \simeq 0.5$), the binary's orbital eccentricity generally does not reach a very large value; 
therefore, the collision fraction is large for very small $\alpha$ values ($\alpha \simeq 2$), but decreases rapidly as $\alpha$ increases.
For larger $\beta_b$ (e.g., $0.8$), the binary's orbital eccentricity can reach a higher value after the close encounter, but the binary disruption occurs more frequently, causing the stars to separate before they can collide. 
In this case, the collision fraction is larger for large $\alpha$ values ($\alpha \simeq 15$) and depends more weakly on $\alpha$.

The results of gentle encounters can be qualitatively understood by using single-averaged secular equations for triples.
The characteristic pericenter passage time is $t_{\rm peri} \sim \sqrt{r_p^3/(GM)}$, and the orbital timescale is $t_{\rm orb} \sim \sqrt{a_b^3/(Gm_{12})}$, and the ratio is
\begin{equation}
    \frac{t_{\rm peri}}{t_{\rm orb}}\sim \Big( \frac{r_p}{a_b} \Bigr)^{3/2}\Big( \frac{m_{12}}{M} \Bigr)^{1/2} = \beta_b^{-3/2}.
\label{eq:periods}
\end{equation}
For $t_{\rm peri}/t_{\rm orb} \gg 1$ (or $\beta_b \ll 1$), the internal orbital motion of the binary can be averaged across the entire encounter, and single-averaged secular equations of motion can be obtained.
To the leading (quadruple) order, we have [see Eqs.~(28), (29) in \citealt{Liu2018ApJ}]:

\begin{align}
        \frac{d{\boldsymbol j}}{dt} &= \frac{3}{2t_{\rm LK}}[5(\boldsymbol{e} \cdot     \hat{\boldsymbol{r}}_{\rm CM}) \boldsymbol{e} \times \hat{\boldsymbol{r}}_{\rm CM}-(\boldsymbol{j} \cdot \hat{\boldsymbol{r}}_{\rm CM}) \boldsymbol{j} \times \hat{\boldsymbol{r}}_{\rm CM}], \label{eq:sa_j} \\
        \frac{d{\boldsymbol e}}{dt} &= \frac{3}{2t_{\rm LK}}[5(\boldsymbol{e} \cdot     \hat{\boldsymbol{r}}_{\rm CM}) \boldsymbol{j} \times \hat{\boldsymbol{r}}_{\rm CM}-(\boldsymbol{j} \cdot \hat{\boldsymbol{r}}_{\rm CM}) \boldsymbol{e} \times \hat{\boldsymbol{r}}_{\rm CM} \notag \\
        & -2 \boldsymbol{j} \times \boldsymbol{e}], \label{eq:sa_e}
\end{align}
where $\boldsymbol{e} = e\hat{\boldsymbol{e}}$ points in the direction of the pericenter, 
$\boldsymbol{j} = \sqrt{1-e^2}\hat{\boldsymbol{L}}$ points in the direction of the binary orbital angular momentum,
$t_{\rm LK}$ is the Lidov-Kozai timescale defined as:
\begin{equation}
    t_{\rm LK}\equiv\frac{m_{12}}{M}\Bigl( \frac{r_{\rm CM}}{a_b} \Bigr)^3 \Bigl( \frac{a_b^3}{Gm_{12}} \Bigr)^{1/2}.
\end{equation}

\begin{figure}[htbp]
\centering
\includegraphics[width=1.02\columnwidth]{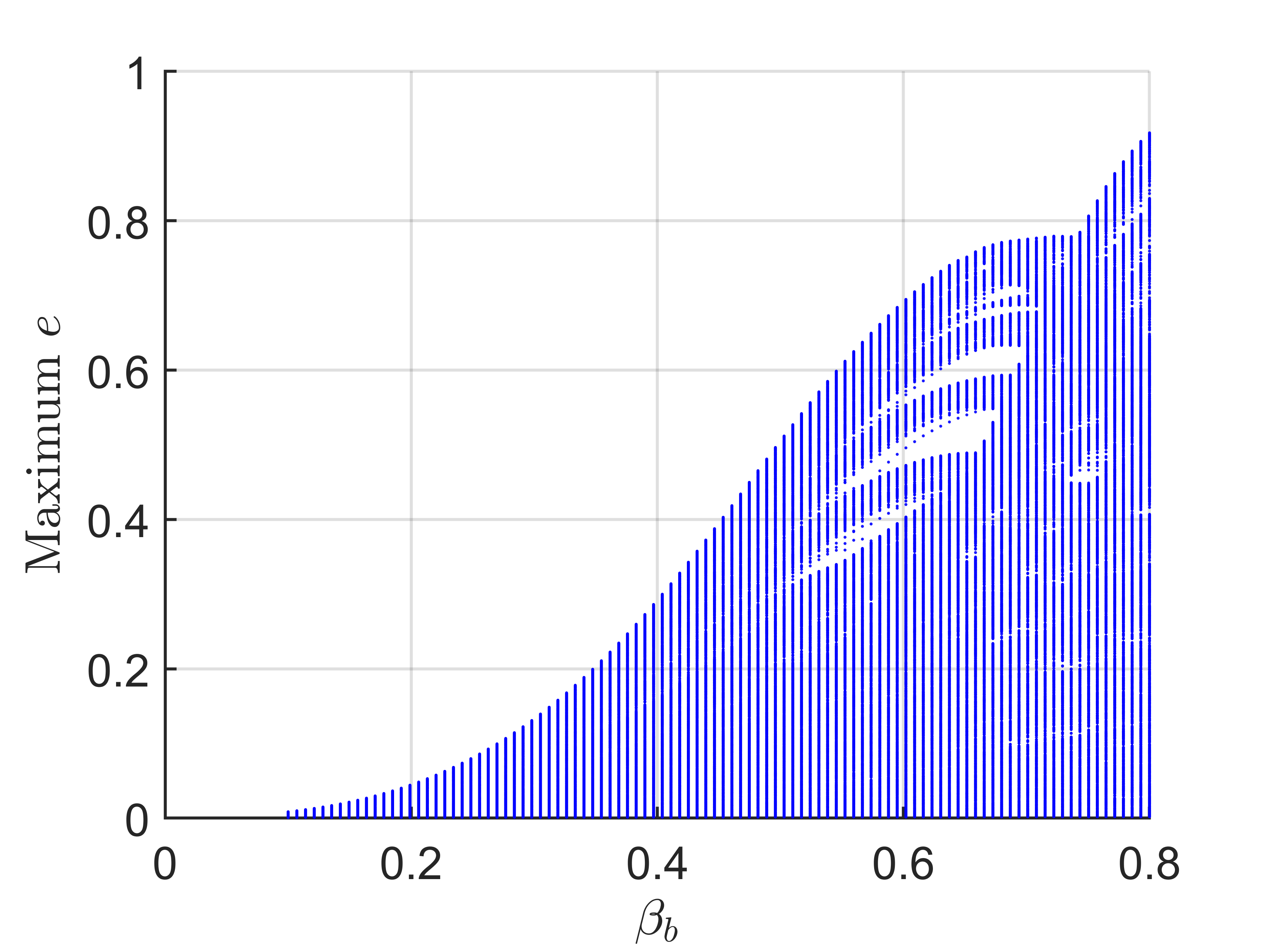}
\caption{
The maximum eccentricity the binary can achieve during gentle binary-SMBH encounters as a function of $\beta_b$, calculated using the single-averaged secular equations (\ref{eq:sa_j})-(\ref{eq:sa_e}).
Each blue dot represents the result of one calculation (with a specific set of initial conditions). 
Note that the gaps in the middle range of $\beta_b$'s result from the finite samplings of the initial angles and phases.}
\label{Fig:Eexcitation}
\end{figure}

For a given $\beta_b$, we substitute Eqs.~(\ref{eq:rCM})-(\ref{eq:ttof}) into Eqs.~(\ref{eq:sa_j})-(\ref{eq:sa_e}), numerically integrate them for various initial phase angles and inclinations, and record the maximum eccentricity of the binary throughout the whole encounter. 
The result is shown in Figure~\ref{Fig:Eexcitation}.
There is a clear envelope for the maximum eccentricities.
Stellar collision occurs when $a_b(1-e)\leq R_{\rm col}$, or $1-e \leq \alpha^{-1}$.

We see that as $\beta_b$ increases, the binary can attain a significant eccentricity, making stellar collision possible after the encounter. 
This is in agreement with the $N$-body simulation results shown in Figure~\ref{Fig:col_frac_lowbeta}.

\subsubsection{Deep Encounters (\texorpdfstring{$\beta_b \gtrsim 1$}{β>1})}
\label{sec:br_regime2}

During a deep encounter, when the center of mass of the binary reaches $r_{\rm tide}^b$, the binary is disrupted energy-wise. 
After this point, the motion of the two stars becomes almost independent. 
Since the velocities of the two stars differ slightly when $r_{\rm CM}$ reaches $r_{\rm tide}^b$, their trajectories also differ, potentially leading to a collision around the pericenter. 
Such collisions are relatively rare, but there is the possibility of high collision velocities.

\begin{figure}[htbp]
\centering
\includegraphics[width=1.02\columnwidth]{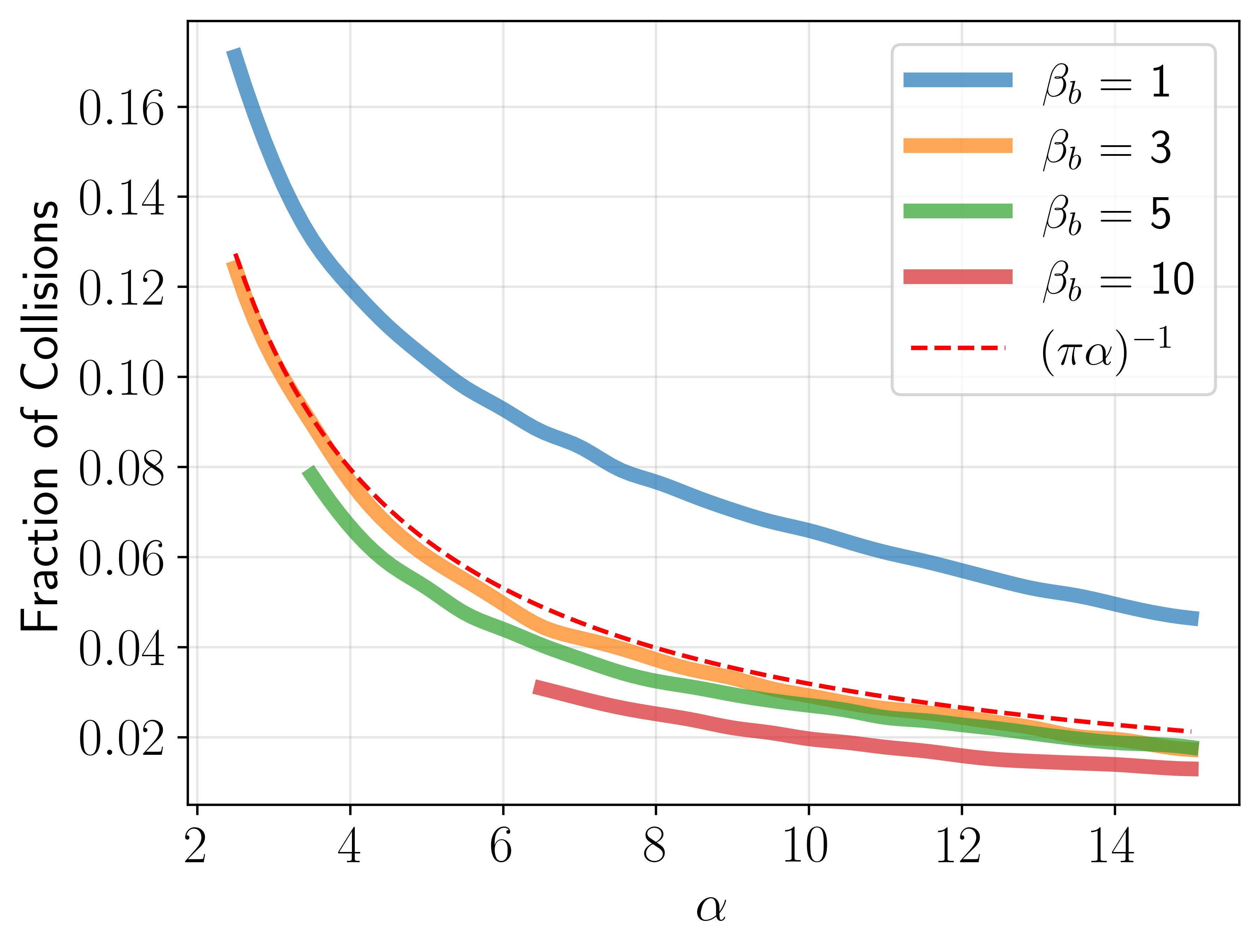}
\caption{
Same as Figure~\ref{Fig:col_frac_lowbeta}, except for deep encounters with $\beta_b \gtrsim 1$.
The solid lines correspond to $\beta_b=1, 3, 5$ and $10$.
The red dashed line represents the ``billiard ball" model in which the collision fraction is $1/(\pi \alpha)$.}
\label{Fig:col_frac_highbeta}
\end{figure}

Figure~\ref{Fig:col_frac_highbeta} shows the fraction of stellar collisions as a function of $\alpha$, for simulations with $\beta_b =1, 3, 5$ and $10$.
We see that the fraction is smaller (by a factor of a few) than the case of gentle encounters (see Figure~\ref{Fig:col_frac_lowbeta}), and generally decreases with increasing $\alpha$ and $\beta_b$.
For such deep encounters, a simple ``billiard ball" model discussed by \citet{Ginsburg2007MNRAS} provides a rough estimate of the collision fraction: 
when the interaction between the two stars is neglected, the direction of the relative velocity between the stars is completely random; 
a stellar collision occurs when the angle $\Delta \theta$ between the relative velocity vector and the relative position vector is less than $2R_{\rm col}/a_b$; 
thus the probability of a collision is $\Delta \theta/ 2\pi= 1/(\pi \alpha)$.
We see from Figure~\ref{Fig:col_frac_highbeta} that this provides a very good fit to the $\beta_b = 3$ result
\citep[see also Table.~1 of][]{Ginsburg2007MNRAS}.

\subsubsection{Eccentric Binaries}

For completeness, here we examine the impact of the initial binary eccentricity on the collision fraction. 
Note that to avoid stellar collision before the binary-SMBH encounter, we require $a_b(1-e_b)>R_{\rm col}$, or $\alpha>1/(1-e_b)$.

\begin{figure}[htbp]
        \subfigure
        {
            \begin{minipage}[b]{0.95\linewidth} 
                \centering
                \includegraphics[width=\columnwidth]{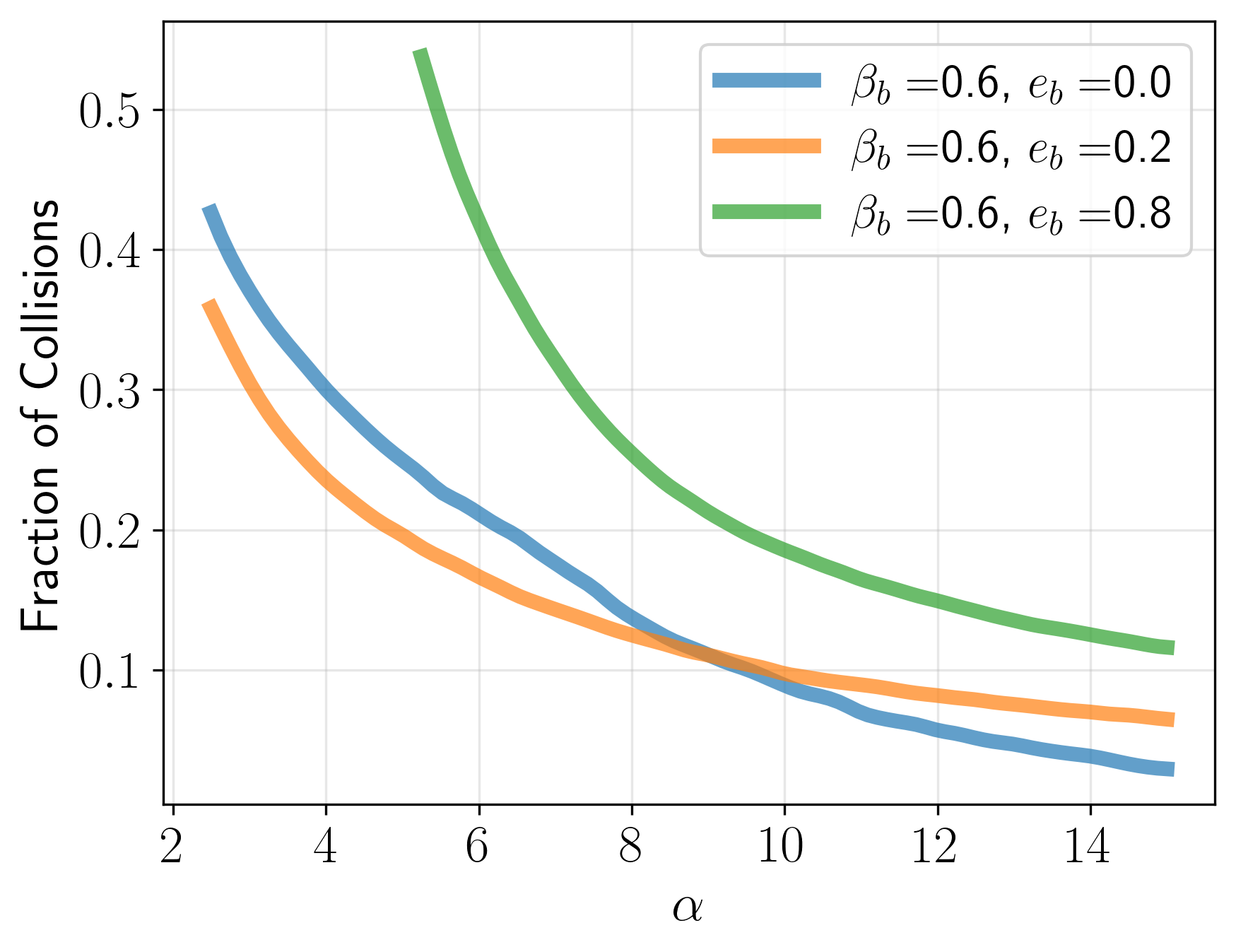}
            \end{minipage}
        }
        \subfigure
        {
            \begin{minipage}[b]{0.95\linewidth}
                \centering
                \includegraphics[width=\columnwidth]{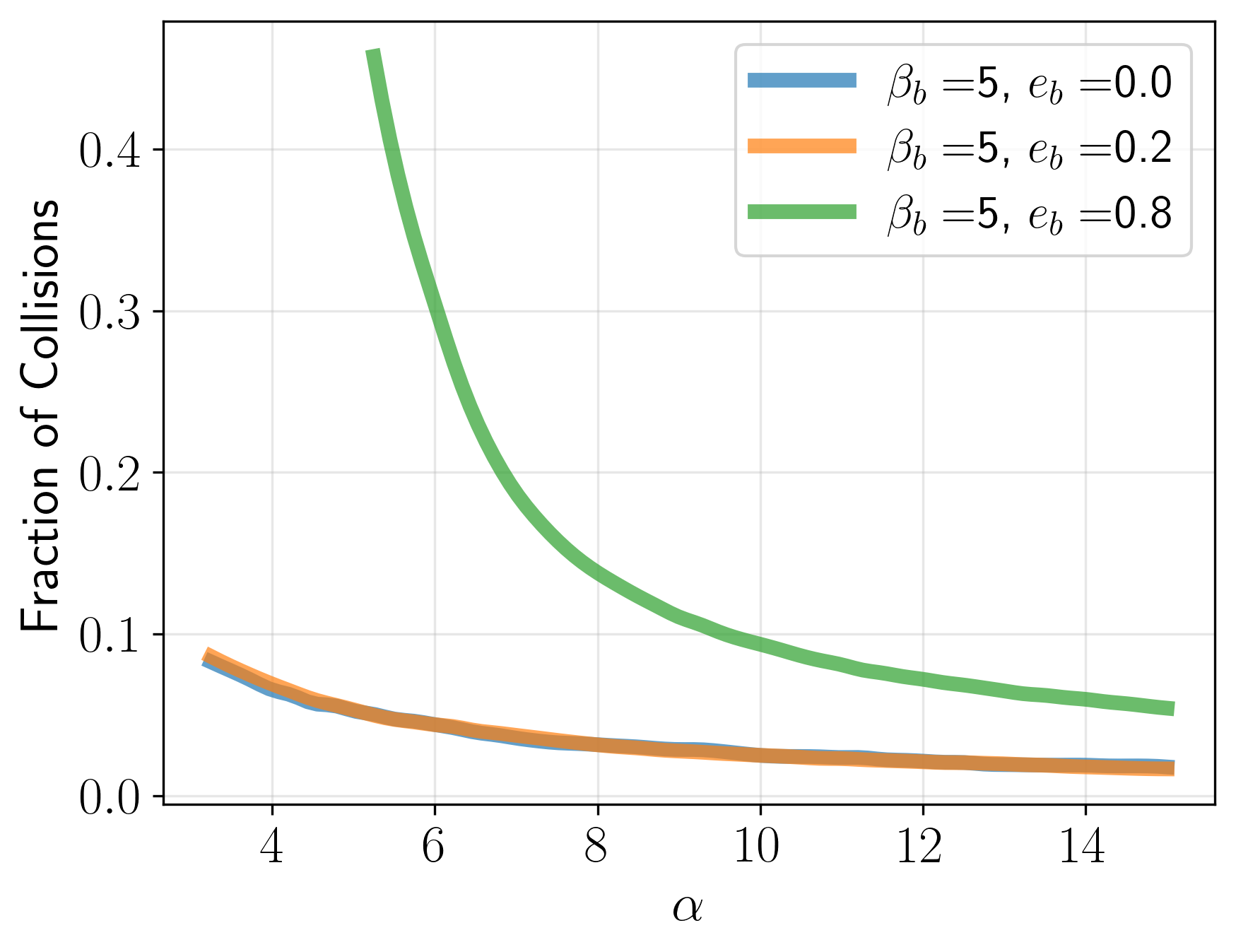}
            \end{minipage}
        }
        \caption{
        Same as Figs.~\ref{Fig:col_frac_lowbeta} and \ref{Fig:col_frac_highbeta}, except for different initial binary eccentricities with two selected $\beta_b$.}
        \label{Fig:Diffe_col_frac}
\end{figure}

Figure~\ref{Fig:Diffe_col_frac} illustrates the effect of different initial binary eccentricities on the collision fraction in both gentle encounters ($\beta_b=0.6$) and deep encounters ($\beta_b=5$). We see that for $e_b\lesssim 0.2$, the collision fraction is similar to the $e_b=0$ case. 
However, for higher eccentricities ($e_b\sim 0.8$), the collision fraction can be much larger. 
This suggests that if multiple encounters occur between the binary and the SMBH before tidal circularization can sufficiently reduce the eccentricity, the fraction of binary collisions will increase significantly \citep[see][]{Antonini2010ApJ, Bradnick2017MNRAS}.

\subsection{Collision Velocity}
\label{sec:collision velocity}

When two stars collide, it is important to know their relative velocity and impact parameter, as they determine the outcome of the collision including the mass loss \citep[e.g.,][]{Benz1987ApJ,Lai1993ApJ,Freitag2005MNRAS}.

\begin{figure}[htbp]
        \subfigure
        {
            \begin{minipage}[b]{0.95\linewidth} 
                \centering
                \includegraphics[width=\columnwidth]{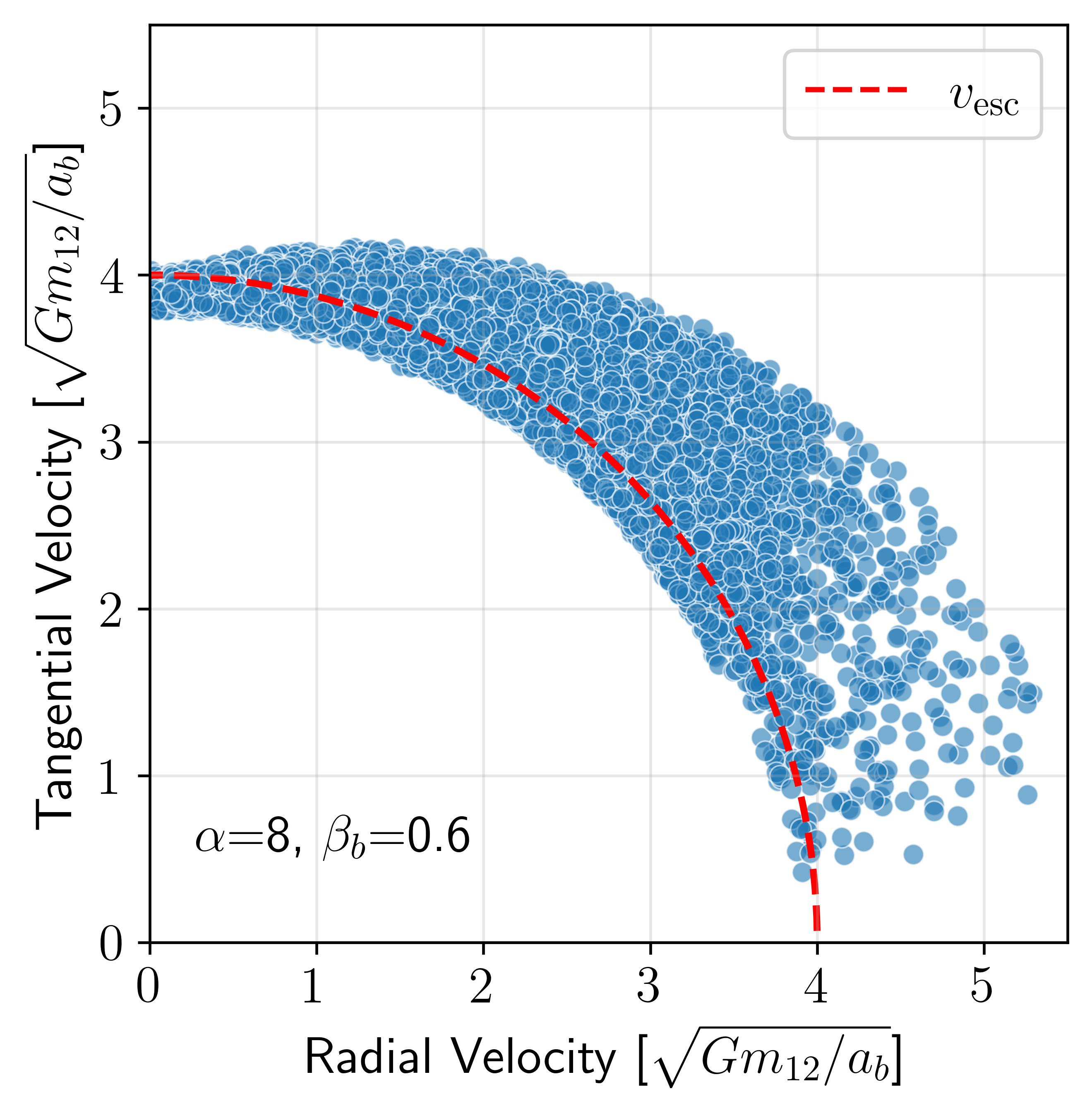}
            \end{minipage}
        }
        \subfigure
        {
            \begin{minipage}[b]{0.95\linewidth}
                \centering
                \includegraphics[width=\columnwidth]{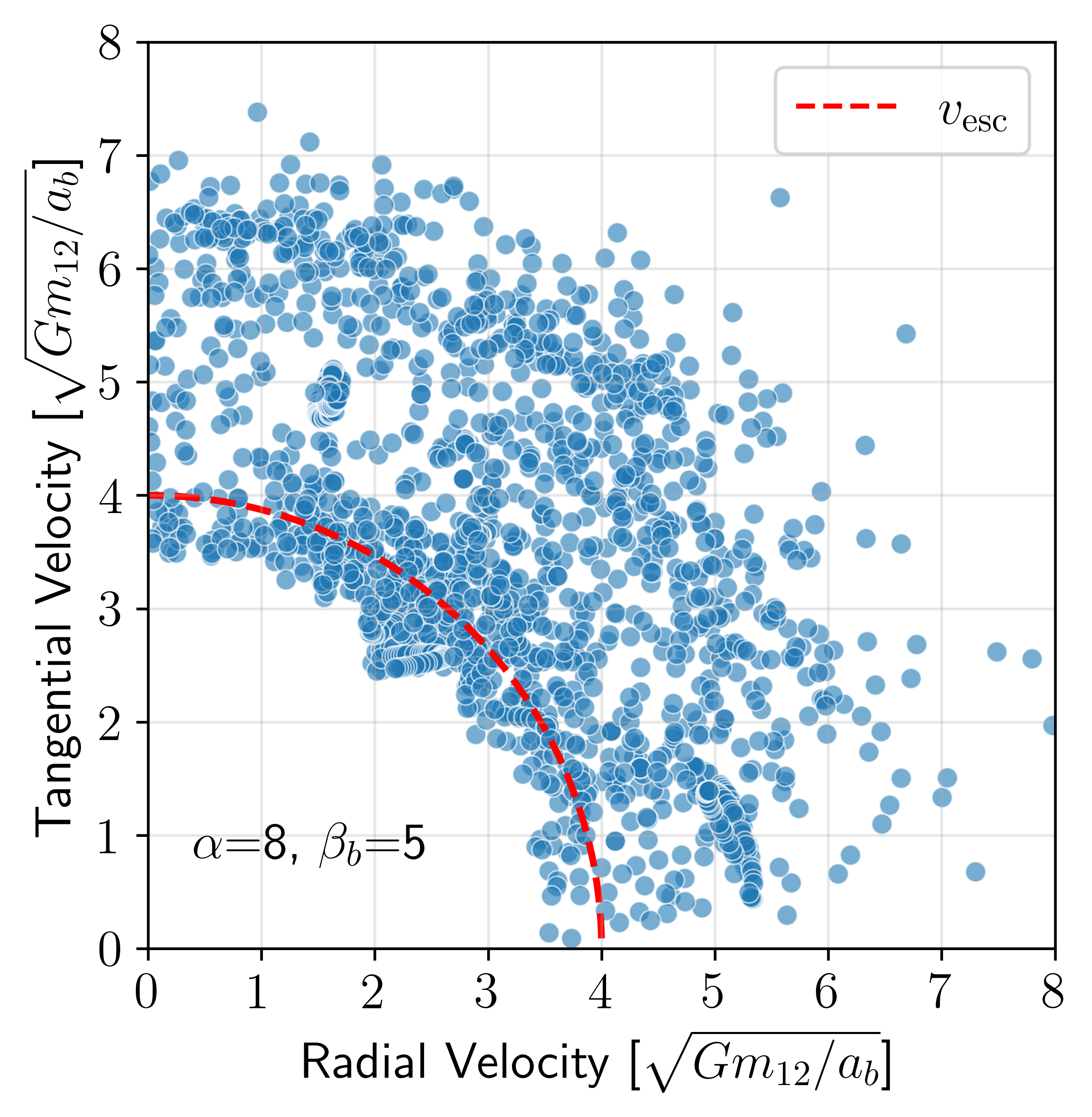}
            \end{minipage}
        }
        \caption{
        Distribution of the relative velocities between the two stars at collision for $\beta_b=0.6$ (upper panel) and $\beta_b=5$ (lower panel). 
        The horizontal axis represents the radial velocity $v_{\rm col,r}$ and the vertical axis represents the tangential velocity $v_{\rm col,t}$.
        The red dashed line indicates the escape velocity $v_{\rm esc}=\sqrt{2Gm_{12}/R_{\rm col}}$.
        Each blue dot in the figure represents the result of one simulation.}
        \label{Fig:collision_v}
\end{figure}

Figure~\ref{Fig:collision_v} shows the distribution of collision velocities for two typical $\beta_b$ values ($\beta_b=0.6$ for gentle encounters and $\beta_b=5$ for deep encounters), both with $\alpha = 8$.
Note that these results can be scaled to different $\alpha$ values: 
if we know that the same orbital parameters can lead to collisions for a larger $\alpha$ value, we can analytically calculate the new collision velocity components, since the interaction between the two stars dominates the relative motion at close distances.

We see from Figure~\ref{Fig:collision_v} that the collision velocity $v_{\rm col}$ ranges from slightly less than the escape velocity $v_{\rm esc}$ to twice the escape velocity, where $v_{\rm esc} \equiv (2Gm_{12}/R_{\rm col})^{1/2}$. [Note that for two identical stars, $v_{\rm esc} = (2Gm/R_{\star})^{1/2}$]
This is not surprising, since for $\alpha \gg 1$, the collision velocity should be exactly equal to $v_{\rm esc}$.

In Figure~\ref{Fig:collision_v}, large tangential velocity ($v_{\rm col,t} \gg v_{\rm col,r}$) implies nearly grazing collisions, while large radial velocity ($v_{\rm col,r} \gg v_{\rm col,t}$) implies nearly head-on collisions. 
We see that both head-on and grazing collisions are possible.
Figure \ref{Fig:collision_v} also shows that there is a significant difference in the collision velocities for the two $\beta_b$ values.
For gentle encounters (upper panel), the maximum collision velocity is only slightly higher than the escape velocity, and collisions tend to be more grazing than head-on.
For deep encounters (lower panel), the maximum collision velocity can reach twice the escape velocity. 
Such collisions will result in more mass loss or even complete destruction of the star, potentially leading to observable effects.

\subsection{Properties of Merger Remnants}
\label{sec:properties of merger remnant}

Although the analysis of collision velocities suggests that some mass loss is inevitable, most collisions will result in a stellar merger. 
Here we examine the properties of merger remnants. 
Without detailed information from hydrodynamics simulations, we will use the simplest model, i.e., assume that the collision is perfectly inelastic, conserving momentum with no mass loss. 

We discuss the orbital energy distribution and
orbital periods of merger remnants relative to the SMBH in Section~\ref{sec:property_energy},
and the fraction of remnants captured by (i.e., bound to) the SMBH in Section~\ref{sec:property_capture}.

\subsubsection{Orbital Energy Distribution}
\label{sec:property_energy}

\begin{figure}[htbp]
        \subfigure
        {
            \begin{minipage}[b]{0.95\linewidth} 
                \centering
                \includegraphics[width=\columnwidth]{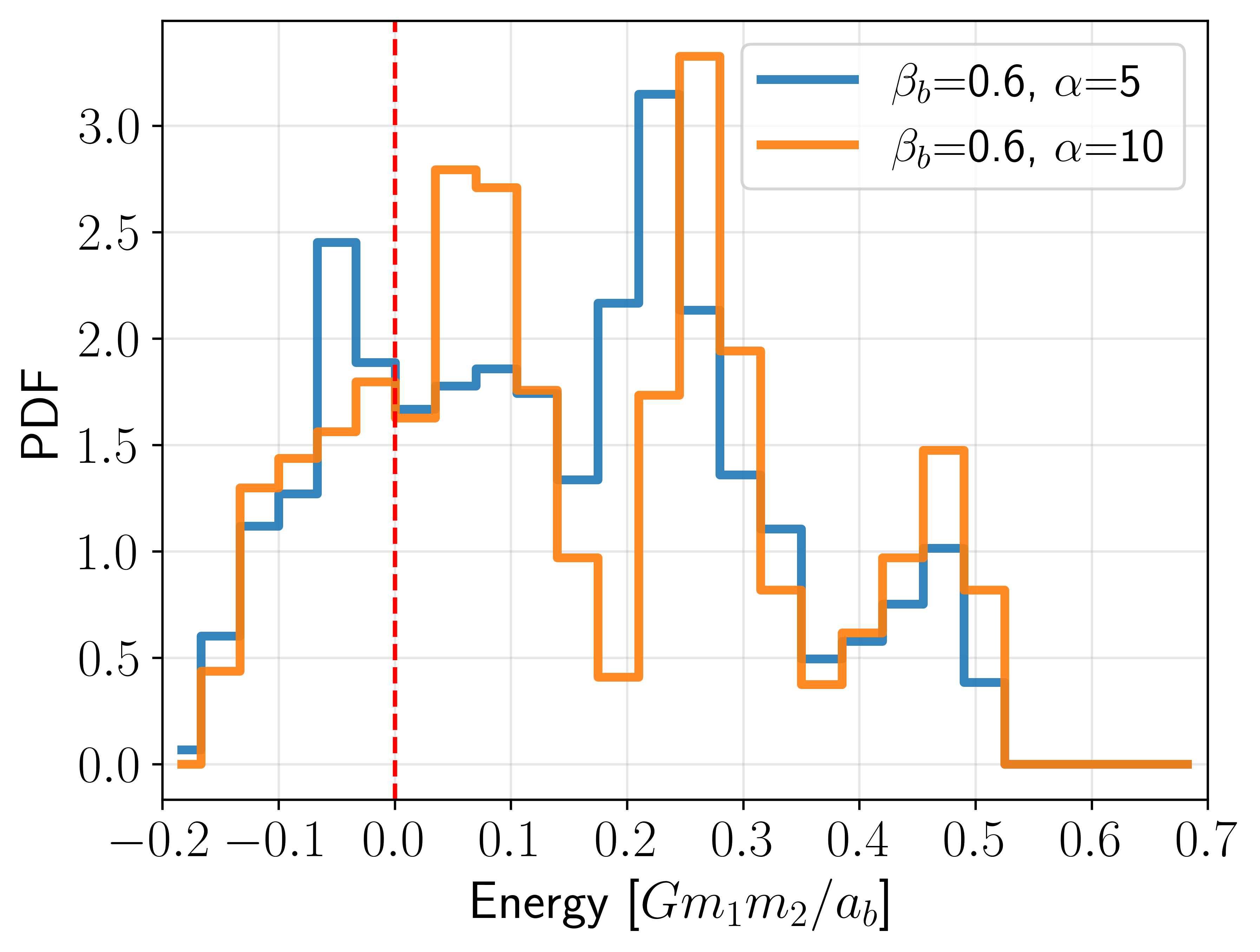}
            \end{minipage}
        }
        \subfigure
        {
            \begin{minipage}[b]{0.95\linewidth}
                \centering
                \includegraphics[width=\columnwidth]{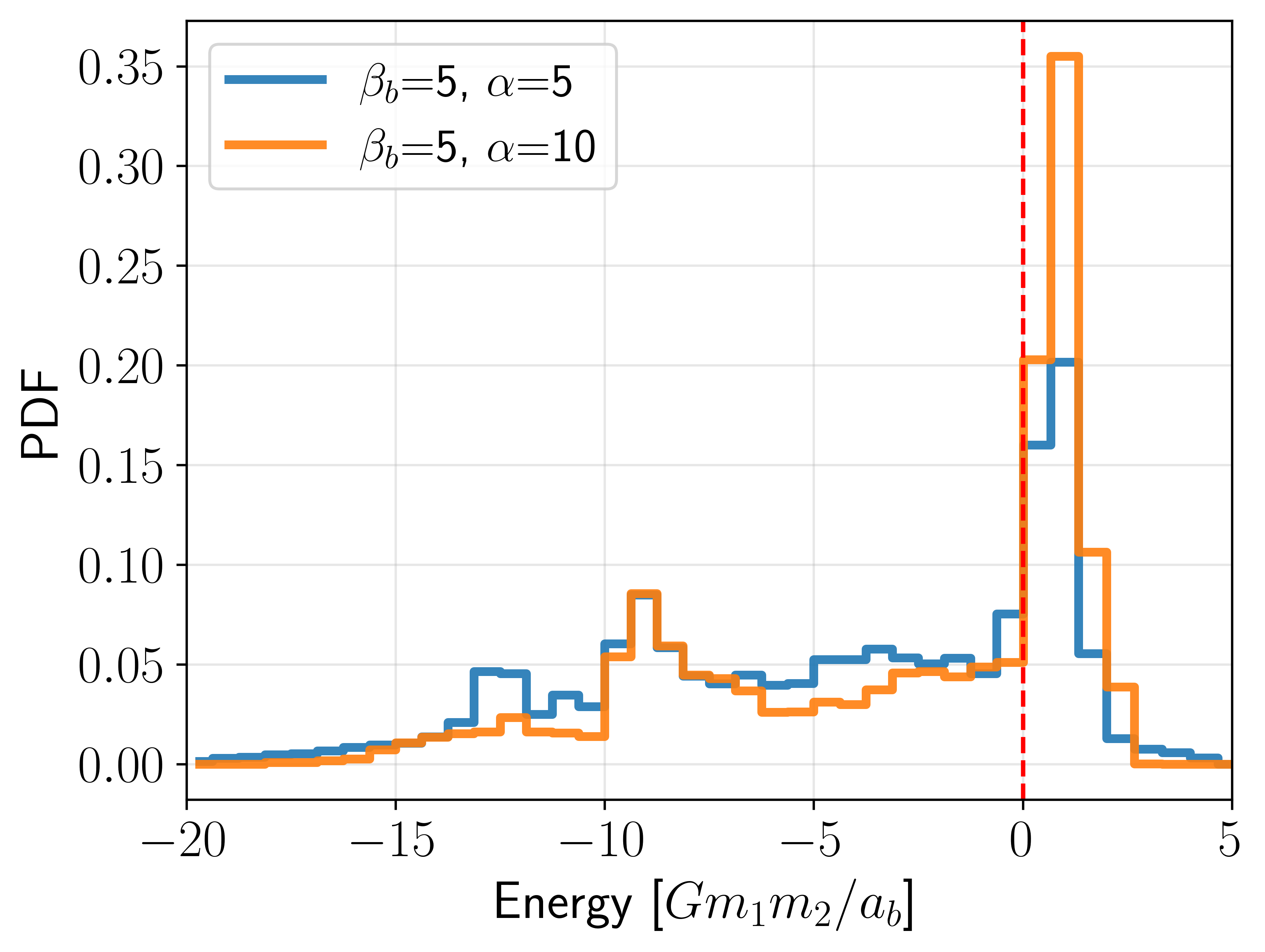}
            \end{minipage}
        }
        \caption{
        Probability density function of the remnant orbital energy for the gentle encounters ($\beta_b=0.6$, upper panel) and deep encounters ($\beta_b=5$, lower panel), each with two values of $\alpha$.
        The vertical red dashed line is the dividing line that distinguishes whether a remnant is captured by the SMBH or not.}
        \label{Fig:remnant_E}
\end{figure}

Since the center of mass of the binary is on a nearly parabolic orbit, its initial orbital energy should be zero. 
Following the inelastic stellar collision, the resulting merger remnant will exhibit a distribution of orbital energies.
Figure \ref{Fig:remnant_E} shows the energies of the remnants for both gentle encounters ($\beta_b = 0.6$) and deep encounters ($\beta_b = 5$), with $\alpha = 5, 10$.

The difference in the energy distributions between the gentle and deep encounters is significant.
For gentle encounters, the absolute value of the energy $E_{\rm rem}$ is small, of order 
\begin{equation}
   |E_{\rm rem}|\sim 0.2 \frac{Gm_1m_2}{a_b},
\end{equation}
because collisions occur far from the SMBH. 
As a result, most of the merger remnants are not captured by the SMBH and are instead ejected.
The velocity of the ejected remnant is
\begin{equation}
    v_{\rm ej} \simeq 0.6 \Bigl( \frac{G\mu_{12}}{a_b} \Bigr)^{1/2} \Bigl(\frac{|E_{\rm rem}|}{0.2Gm_1m_2/a_b} \Bigr)^{1/2},
\label{eq:vej}
\end{equation}
where $\mu_{12}=m_1m_2/m_{12}$ is the reduced mass of the binary.
Note that $|E_{\rm rem}|$ is much less than the characteristic tidal energy in binary disruption (Eq.~\ref{eq:Etide}).
Thus the velocity of the ejected remnant (Eq.~\ref{eq:vej}) is much less than the velocity of the ejected star in a binary disruption.
Similarly, the remnants that are captured are only weakly bound to the SMBH.

For deep encounters, collisions generally occur at the pericenter passage, resulting in a much larger $|E_{\rm rem}|$. 
Although the peak of the energy distribution is greater than zero, there is a wide distribution towards negative energies, meaning that most of the merger remnants are captured by (bound to) the SMBH. 
The orbital energy of such a bound remnant can be written as
\begin{equation}
    E_{\rm rem}= -\epsilon_{\rm rem}\frac{Gm_1m_2}{a_b},
   \label{eq:Erem}
\end{equation}
where $\epsilon_{\rm rem}\sim 10$ typically and can be as large as 20.  
Setting $E_{\rm rem}=-GMm_{12}/(2a_{\rm rem})$, this translates into the semi-major axis of the remnant
\begin{equation}
    a_{\rm rem}=-\frac{M}{2\epsilon_{\rm rem}\mu_{12}}a_b,
\label{eq:arem}
\end{equation}
where $\mu_{12}=m_1m_2/m_{12}$, and the orbital period of the remnant 
\begin{equation}
    P_{\rm rem}=\Bigl( \frac{m_{12}}{2\epsilon_{\rm rem}\mu_{12}} \Bigr)^{3/2} \Bigl( \frac{M}{m_{12}}\Bigr) P_b,
\label{eq:Prem_1}
\end{equation}
where $P_b=2\pi \sqrt{a_b^3/(Gm_{12})}$ is the internal binary period.

\begin{figure}[htbp]
 \centering
\includegraphics[width=1.02\columnwidth]{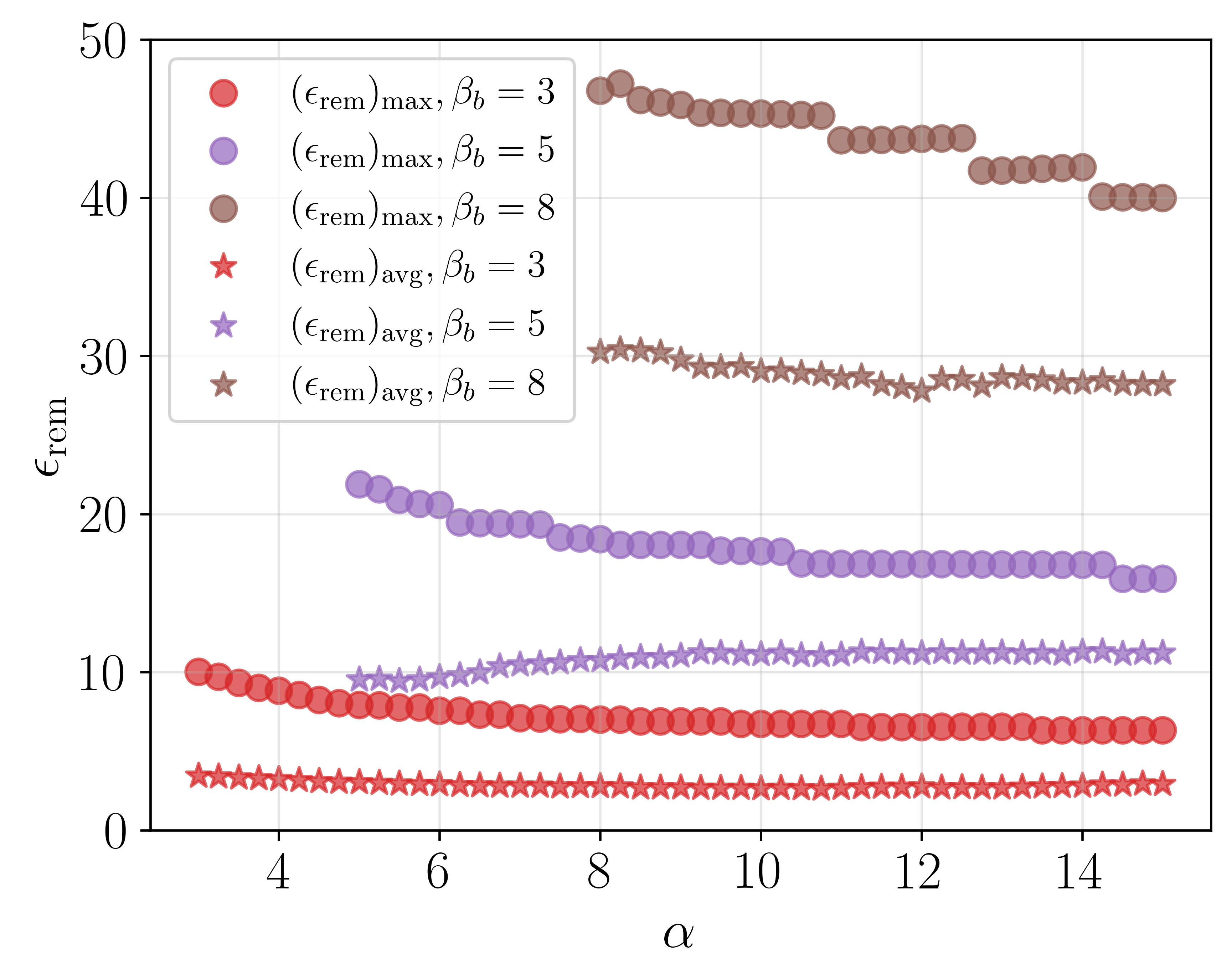}
\caption{
The maximum (point) and average (star) value of the dimensionless energy $\epsilon_{\rm rem}$ (see Eq.~\ref{eq:Erem}) of merger remnants as a function of $\alpha$ for different $\beta_b$.
Each point represents the result of a suite of simulations (about $10^4$ runs).
The jaggedness in the figure comes from numerical precision. } 
\label{Fig:Epsilon}
\end{figure}

Figure~\ref{Fig:Epsilon} shows the relationship of both the largest and average values of $\epsilon_{\rm rem}$ with respect to $\alpha$. 

The results indicate that for deeper encounters (larger $\beta_b$), $\epsilon_{\rm rem}$ is generally larger, and the maximum value gradually decreases as $\alpha$ increases, though the average value shows no clear trend. 
According to Equation~(\ref{eq:Prem_1}), this also suggests that the minimum orbital period decreases when the encounter is deeper.

\subsubsection{Capture Fraction of Merger Remnants}
\label{sec:property_capture}

\begin{figure}[htbp]
        \subfigure
        {
            \begin{minipage}[b]{0.95\linewidth} 
                \centering
                \includegraphics[width=\columnwidth]{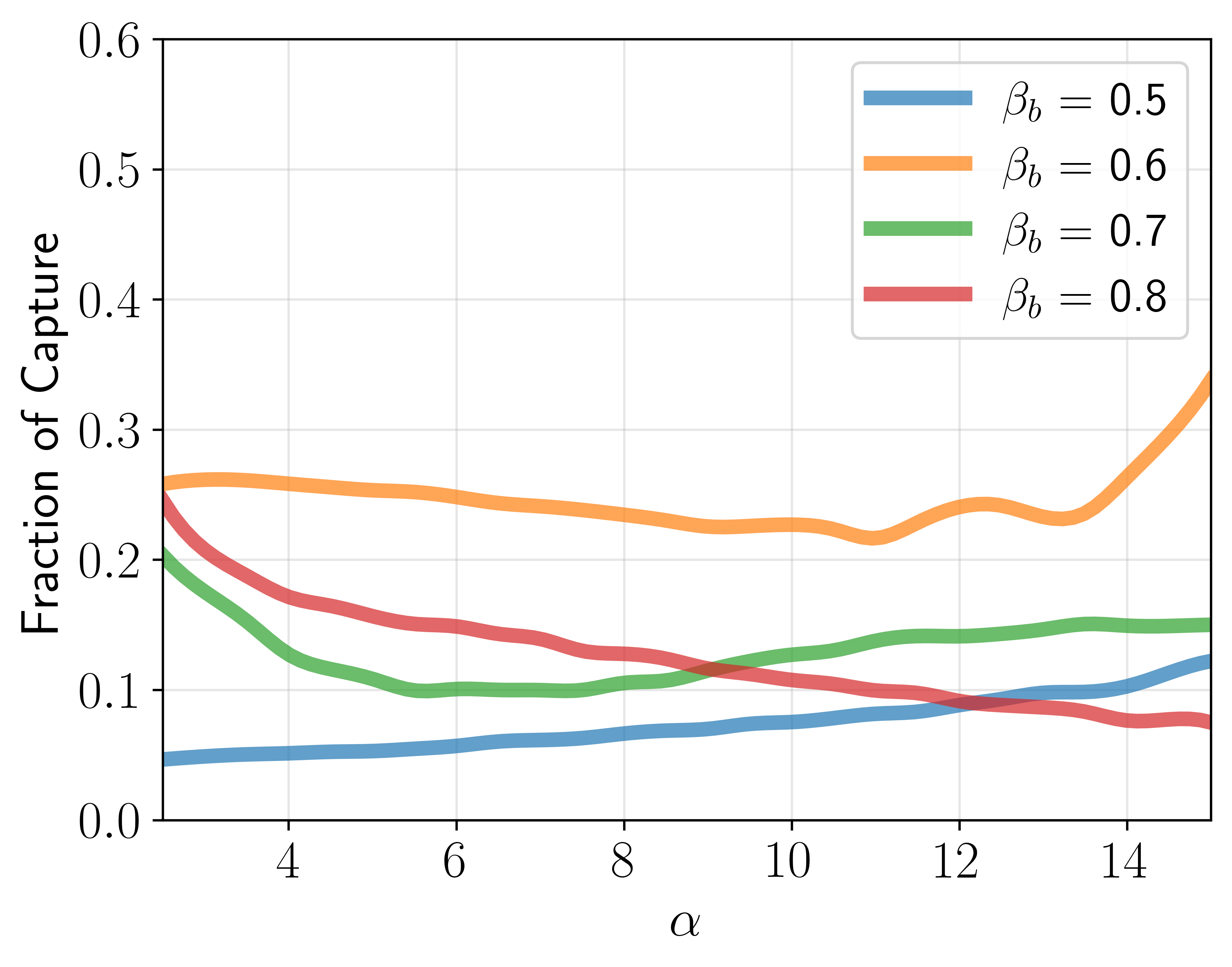}
            \end{minipage}
        }
        \subfigure
        {
            \begin{minipage}[b]{0.95\linewidth}
                \centering
                \includegraphics[width=\columnwidth]{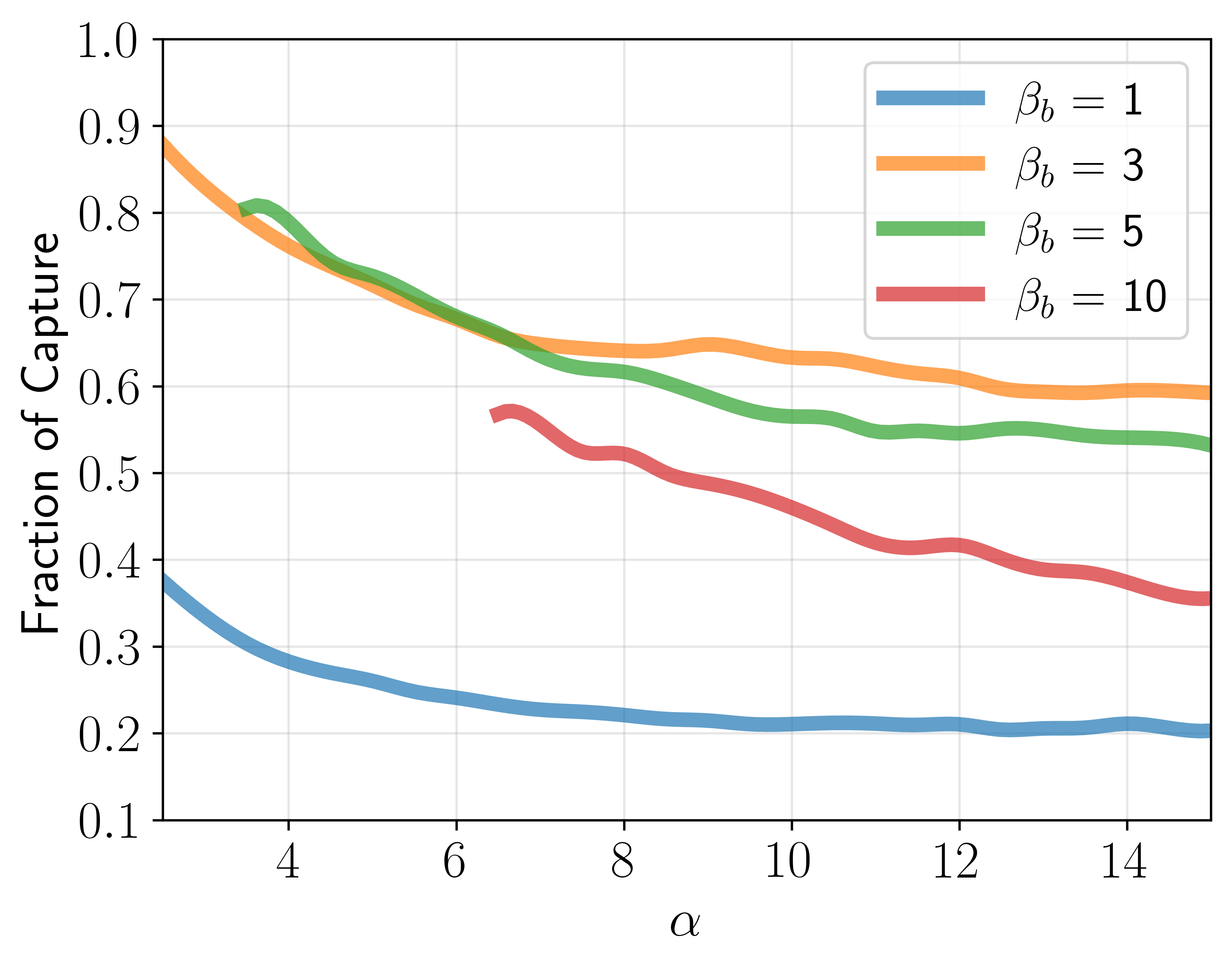}
            \end{minipage}
        }
        \caption{
        Same as Figs.~\ref{Fig:col_frac_lowbeta} and \ref{Fig:col_frac_highbeta}, except that the figure shows the fraction of the stellar merger remnants (among the stellar collisions) that are captured by the SMBH.
        Note that for $\beta_b = 5$ and $10$, we only consider $\alpha > 2^{-2/3}\beta_b$ (see Eq.~\ref{eq:avoidTDE}) to avoid stellar tidal disruption.}
        \label{Fig:capture_fraction}
\end{figure}

We have carried out simulations for a range of $\alpha$ and $\beta_b$ values to determine the fraction of remnants that are captured by the SMBH.
The result is shown in Figure~\ref{Fig:capture_fraction}.
We see that the capture fraction
in gentle encounters is much smaller than in deep encounters.
The result for $\beta_b = 1$ (the blue solid line in the lower panel of Figure~\ref{Fig:capture_fraction}) is different from the $\beta_b > 1$ cases because $\beta_b = 1$ represents a mixture of the two encounter regimes, resulting in a smaller capture fraction compared to the $\beta_b > 1$ cases.

Note that the capture fraction is calculated from the stellar merger events, so $\alpha$ does not affect the capture fraction in the same way it affects the collision fraction in Section~\ref{sec:collision fraction}. 
In gentle encounters, we find that the capture fraction has almost no dependence on $\alpha$. 
In deep encounters, the capture fraction decreases monotonically as $\alpha$ increases. 
A possible explanation is that as $\alpha$ increases, collisions occur farther away from the SMBH and the resulting remnants are less likely to be bound to the SMBH.

\section{Summary and Discussion}
\label{sec:application and summary}

\subsection{Summary of Key Results}
\label{sec:summary}

We have carried out a systematic study on the close encounters between stellar binaries and SMBHs. 
The binary (with a total mass $m_{12}$, initial circular orbit and semi-major axis $a_b$) approaches the SMBH (mass $M$) on a nearly parabolic orbit, with pericenter distance $r_p$ to the SMBH. 
In addition to binary break-ups, such encounters can also lead to double stellar disruptions (double TDEs) and stellar collisions/mergers. 
The occurrences of these different outcomes depend on the system parameters through three dimensionless ratios:
$\beta_b=r_{\rm tide}^b/r_p$, where $r_{\rm tide}^b=a_b(M/m_{12})^{1/3}$ is the tidal radius of the binary; 
$\beta_\star=r_{\rm tide}^\star/r_p$, where $r_{\rm tide}^\star$ is the tidal radius of the star; 
and $\alpha=a_b/R_{\rm col}$, where $R_{\rm col}$ is the separation of the two stars at collision.
(For a binary with identical stars, $R_{\rm col}=2R_\star$ and $\beta_b=2^{2/3}\alpha \beta_\star$.)
Using 3-body scattering experiments, we characterize these different outcomes as a function of the three dimensionless ratios. 

Our main findings concern the stellar collisions in both gentle encounters ($\beta_b\lesssim 1$) and deep encounters ($\beta_b\gtrsim 1$). 
The occurrence fraction of collisions depends on the compactness of the binary (or the parameter $\alpha$), but can reach a few to 10's percent (Figs.~\ref{Fig:col_frac_lowbeta}-\ref{Fig:col_frac_highbeta}).
\begin{itemize}
\item In gentle encounters, the binary survives the pericenter passage, but its orbit becomes very eccentric, leading to a stellar collision with a contact velocity near the escape velocity of the star (the upper panel of Figure~\ref{Fig:collision_v}); 
the merger remnants are typically ejected from the SMBH at a small velocity.
\item In deep encounters, stellar collisions can dynamically occur during the pericenter passage; 
the contact velocity can reach a few times the escape velocity of the star (the lower panel of Figure~\ref{Fig:collision_v}), leading to appreciable mass loss from the collision. 
Most merger remnants are bound to the SMBH, with typical orbital energy of order $-10 Gm_1m_2/a_b$ (see Figure~\ref{Fig:remnant_E} and Equation~\ref{eq:Erem}).
\end{itemize}

Additionally, for completeness, we present the break-up fraction and energy distribution of the stars following a binary breakup for different $\beta_b$'s and initial binary eccentricities (see Figs~\ref{Fig:Diffe_Break_up_frac}-\ref{Fig:Diffe_Hills_E}).
Eccentric binaries are more susceptible to tidal break-up. 
The energy distribution peaks at the characteristic ``tidal" energy (Eq.~\ref{eq:Etide}) for circular and slightly eccentric binaries, but peaks at lower value for highly eccentric binaries, and differs from the energy distribution of the tidal debris produced when a single star is disrupted. 
We also explore the double TDE scenario, where we show that the time interval between successive stellar disruptions is always small, making the difference of the return times of the two disrupted stellar streams more significant.

\subsection{Discussion: Applications of the Results}
\label{sec:application}

Although our main results are presented in dimensionless forms, they can be applied to various systems and situations. 
Here we discuss several potential applications.

\subsubsection{Bound Stars from Binary Break-ups}
\label{sec:app_bound_stars}

It is well-known that the tidal break-up of a binary produces an HVS and a tightly bound star around the SMBH. 
Our calculations show that the energy of the bound star is given by
(see Eq.~\ref{eq:Etide})
\begin{equation}
E_{\rm bs}=-\epsilon_{\rm bs}E_{\rm tide}=-\epsilon_{\rm bs}\frac{GMm a_b}{(r_{\rm tide}^b)^2},
\end{equation}
where for simplicity we have assumed $m_1=m_2=m$ (so that $\alpha=a_b/2R_\star$).
The coefficient $\epsilon_{\rm bs}$ peaks at around 1.2, and can extend to $\sim 2$ for large $\beta_b$ (see Figure~\ref{Fig:Hills_Edistribution}).
Such a bound star has a semi-major axis and eccentricity given by
\begin{eqnarray}
  && a_{\rm bs}=\frac{a_b}{2\epsilon_{\rm bs}} \Bigl(\frac{M}{2m}\Bigr)^{2/3},\\
  && 1-e_{\rm bs}=\frac{r_p}{a_{\rm bs}}=\frac{2\epsilon_{\rm bs}}{\beta_b}
  \Bigl(\frac{2m}{M}\Bigr)^{1/3}\ll 1.
\end{eqnarray}  
The orbital period is 
\begin{eqnarray}
  &&  P_{\rm bs}=\frac{1}{4\epsilon_{\rm bs}^{3/2}}
 \!  \left(\frac{M}{m}\right)^{\!1/2}\!  P_b\nonumber\\
&& =58
  \left(\!\frac{\alpha}{\epsilon_{\rm bs}}\!\right)^{\!\!3/2}
\!\!  \left(\!\frac{M}{10^6m}\!\right)^{\!1/2}
\!\!  \left(\!\frac{R_\star}{R_\odot}\!\right)^{\!3/2}
\!\!  \left(\!\frac{m}{M_\odot}\!\right)^{\!-1/2}\!{\rm days},
\label{eq:Pbs}\end{eqnarray}
where $P_b$ is the initial binary orbital period.
Thus $P_{\rm bs}$ depends sensitively on $a_b/\epsilon_{\rm bs}=2\alpha R_\star/\epsilon_{\rm bs}$.

Bound stars from binary break-ups have been invoked to explain the so-called repeated partial TDEs. 
Several candidate events have been found/claimed recently, including 
ASASSN-14ko 
\citep[period 115.2 days;][]{Payne2021ApJ,Payne2022ApJ,Payne2023ApJ,Huang2023ApJL}, 
eRASSt J045650.3–203750 
\citep[$299 \rightarrow 193$ days;][]{Liu2023AA,Liu2024AA}, 
AT2018fyk 
\citep[$\sim 1200$ days;][]{Wevers2023ApJL}, 
RX J133157.6–324319.7 
\citep[$\sim 10000$ days;][]{Hampel2022RAA,Malyali2023MNRAS}, 
AT 2020vdq 
\citep[$\sim 870$ days;][]{Somalwar2023arXiv}
and AT 2022dbl 
\citep[$\sim 710$ days;][]{Lin2024ApJL}.  
However, in many cases, this interpretation should be considered tentative.  
Since $\alpha$ is greater than a few for realistic binaries, most binary break-ups would not lead to immediate stellar disruption, unless $r_p\lesssim r_{\rm tide}^\star$, or $\alpha/\beta_b \lesssim 1$.  
Equation (\ref{eq:Pbs}) gives the orbital period of the bound star immediately after the binary break-up. 
If the orbit is in the regime where the two-body scattering time is shorter than the gravitational radiation timescale, the eccentricity may continue to increase, potentially leading to a TDE.
On the other hand, if the gravitational radiation timescale is shorter, the orbit will decay \citep{Sari2019ApJ}, and evolve according to
\citep{Peters1964PR}
\begin{equation}
  a(1-e)\propto {e^{12/19} \over 1+e} \left(1+{121e^2\over 304}\right)^{870/2299}.
\end{equation}
So a reduction of the pericenter distance by a factor of a few would imply a reduction of the eccentricity to $\lesssim 0.2$. 
A TDE with such a low eccentricity would be different from the standard high-eccentricity TDEs.

\subsubsection{Merger Remnants from Stellar Collsions}
\label{sec:app_merger_remnants}

Our finding that an appreciable fraction of binary-SMBH encounters lead to stellar collisions may have interesting implications.

As discussed in Section \ref{sec:br_regime1}, gentle encounters ($\beta_b\lesssim 1$) result in mild stellar collisions, and the merger remnants are ejected from the SMBH with small velocities. 
These ejected merger remnants typically have rapid rotations (because of the off-centered collisions) and may contribute to the exotic stellar populations at the centers of galaxies.

On the other hand, for deep encounters, the merger remnants are likely bound to the SMBH, with semi-major axis and eccentricity given by (see Eq.~\ref{eq:arem}, assuming $m_1=m_2=m$)
\begin{eqnarray}
  && a_{\rm rem}=\frac{a_b}{\epsilon_{\rm rem}} \Bigl(\frac{M}{m}\Bigr),\\
  && 1-e_{\rm rem}=\frac{r_p}{a_{\rm rem}}=\frac{\epsilon_{\rm rem}}{2^{1/3}\beta_b}
  \Bigl(\frac{m}{M}\Bigr)^{2/3}\ll 1.
\end{eqnarray}  
The orbital period is 
\begin{eqnarray}
&& P_{\rm rem}=\frac{\sqrt{2}}{\epsilon_{\rm rem}^{\!3/2}}
  \!  \left(\frac{M}{m}\right)P_b\nonumber\\
  &&= 10\left(\!\frac{20
    \alpha}{\epsilon_{\rm rem}}\!\right)^{\!\!3/2}
\!\!  \left(\!\frac{M}{10^6m}\!\right)
\!  \left(\!\frac{R_\star}{R_\odot}\!\right)^{\!3/2}
\!\!  \left(\!\frac{m}{M_\odot}\!\right)^{\!-1/2}\!{\rm years}.
\label{eq:Prem_2}
\end{eqnarray}
(Note that $\epsilon_{\rm rem}$ can be much larger than unity; see Figure~\ref{Fig:Epsilon}.)
Comparing to Equation~(\ref{eq:Pbs}), we see that typically $P_{\rm rem}\gg P_{\rm bs}$.
Thus these merger remnants could also provide an explanation for the repeated partial TDEs, especially for smaller $M/m$ or/and low-mass stars.
One advantage of the merger remnant scenario is that because of dissipation in the merger process, the merger remnant stars likely have a very extended envelope. 
This makes their partial disruptions more likely even without significant orbital decay due to gravitational radiation.
\citet{Bradnick2017MNRAS} have suggested that stellar mergers can produce strong magnetic fields, and the disruption of such stars may lead to jetted TDEs.

Another by-product of stellar collisions induced by binary-SMBH encounters is the mass loss.
Some of this mass will accrete onto the SMBH, generating accretion luminosity/flare even without the TDE of the star. We plan to examine these hydrodynamical processes in future works.

\vspace{0.4cm}
We are grateful to Re'em Sari for useful discussions, and the anonymous referee for constructive comments and suggestions that greatly improved the manuscript.


\software{
IAS15 \citep{reboundias15},
Matplotlib \citep{Hunter2007}, 
NumPy \citep{Walt2011},
Rebound \citep{rebound}.
}

\bibliography{sample631}{}
\bibliographystyle{aasjournal}

\end{document}